\documentclass[%
 aip,
 amsmath,amssymb,
 reprint,%
nofootinbib]{revtex4-1}

\usepackage{graphicx}
\usepackage{dcolumn}
\usepackage{bm}
\usepackage[usenames]{color}
\usepackage[dvipsnames]{xcolor}
\usepackage[utf8]{inputenc}
\usepackage[T1]{fontenc}
\usepackage{mathptmx}

\usepackage{soul}
\usepackage[normalem]{ulem} 
\usepackage{ulem} 

\newcommand{\HC}[1]{\textcolor{black}{cluster}} 
\newcommand{\plural}[1]{\textcolor{black}{clusters}} 
\begin{document}

\preprint{AIP/123-QED}

\title{Mobile oscillators in a mobile multi-cluster network}

\author{Venceslas Nguefoue Meli}
 \affiliation{Research Unit Condensed Matter, Electronics and Signal Processing,
University of Dschang, P.O. Box 67 Dschang, Cameroon.}
\affiliation{MoCLiS Research Group,   P.O. Box 67 Dschang, Cameroon}

\author{Thierry Njougouo}%
\affiliation{IMT School for Advanced Studies Lucca,  Piazza San Francesco 19, 55100 Lucca, Italy}
\affiliation{Faculty of Computer Science and Namur Institute for Complex Systems, naXys Université de Namur,
Rue Grandgagnage 21, B5000 Namur, Belgium}%
\affiliation{MoCLiS Research Group,  P.O. Box 67 Dschang, Cameroon}

\author{Steve J. Kongni }
 \affiliation{Research Unit Condensed Matter, Electronics and Signal Processing,
University of Dschang, P.O. Box 67 Dschang, Cameroon.}
\affiliation{MoCLiS Research Group,  P.O. Box 67 Dschang, Cameroon}

\author{Patrick Louodop}
\affiliation{Research Unit Condensed Matter, Electronics and Signal Processing,
University of Dschang, P.O. Box 67 Dschang, Cameroon.}%
\affiliation{MoCLiS Research Group,  P.O. Box 67 Dschang, Cameroon,}
\affiliation{Potsdam Institute for Climate Impact Research (PIK) Member of the
Leibniz Association P.O. Box 60 12 03 D-14412 Potsdam Germany.}
\affiliation{ICTP South American Institute for Fundamental Research,
S\~ao Paulo State University (UNESP), Instituto de F\'{i}sica Te\'{o}rica,
Rua Dr. Bento Teobaldo Ferraz 271,
Bloco II, Barra Funda, 01140-070 S\~ao Paulo, Brazil.}

\author{Hilaire Bertrand Fotsin}%
\affiliation{Research Unit Condensed Matter, Electronics and Signal Processing,
University of Dschang, P.O. Box 67 Dschang, Cameroon.}%

\author{Hilda A. Cerdeira}
\affiliation{ICTP South American Institute for Fundamental Research,
S\~ao Paulo State University (UNESP), Instituto de F\'{i}sica Te\'{o}rica,
Rua Dr. Bento Teobaldo Ferraz 271,
Bloco II, Barra Funda, 01140-070 S\~ao Paulo, Brazil.}
\affiliation{S\~ao Paulo State University (UNESP), Instituto de F\'{i}sica Te\'{o}rica, Rua Dr. Bento Teobaldo Ferraz 271,
Bloco II, Barra Funda, 01140-070 S\~ao Paulo, Brazil.\\
Epistemic, Gomez $\&$ Gomez Ltda. ME, Rua Paulo Franco 520, Vila Leopoldina, 05305-031 S\~ao Paulo, Brazil\\}

\date{\today}

\begin{abstract}
Different collective behaviors emerging from the unknown have been examined in networks of mobile agents in recent years. Mobile systems, far from being limited to modeling and studying various natural and artificial systems in motion and interaction, offer versatile solutions across various domains, facilitating tasks ranging from navigation and communication to data collection and environmental monitoring. We examine the relative mobility between clusters, each composed of different elements in a multi-clusters network — a system composed of clusters interconnected to form a larger network of mobile oscillators. Each mobile oscillator exhibits both external (i.e., position in a 2D space) and internal dynamics (i.e., phase oscillations). Studying the mutual influence between external and internal dynamics, often leads the system towards a state of synchronization within and between clusters. We show that synchronization between clusters is affected by their spatial closeness. The stability of complete synchronization observed within the clusters is demonstrated through analytical and numerical methods.
\end{abstract}

\maketitle

\begin{quotation}
	 Complex systems, composed of numerous interacting elements, exhibit global behaviors that emerge unpredictably from simple local interactions. These systems, observed in fields as varied as biology, physics, social sciences, and engineering, have garnered increasing attention for their capacity to model complex real-world phenomena \cite{holme2012temporal,fujiwara2011synchronization,majhi2019chimera}. This study explores emergent behaviors in a mobile multi-cluster network of oscillators.Clusters could represent different species where individuals are evolving connecting to each element of the same category while interacting with the global motion of the other group for different reasons, such as foraging success which is the objective of herons staying close to snowy egrets \cite{gloria}, or they could also represent different clusters of the same species. In this work, the analysis focuses on the effects of cluster mobility and stability, followed by the influence of vision range also named vision size--i.e., the maximum  distance beyond which two systems can no longer remain connected. This parameter reflects the decisions made by oscillators to induce spatial deviations in agent positions.
\end{quotation}

\section{Introduction}
The collective motion observed in numerous living systems—such as ant trails, bird flocks, and schools of fish--as well as phenomena like sperm dynamics or sheep migration, has attracted significant interest for its potential to inspire artificial and industrial systems. The concept of networks, which model entities interacting with one another, has provided valuable insights into various fields including biology \cite{ergun2007network, subbalakshmi2022computational, simo2021chimera}, engineering \cite{kitayama2011sequential}, social science \cite{Carrillo2010,couzin2003self,park1927human}, and more. Indeed, phenomena exhibited by collective groups, such as chimera states \cite{abrams2004chimera,zhu2014chimera,majhi2019chimera}, spatial organization or clustering \cite{belykh2008cluster}, and synchronization \cite{belykh2008cluster,pikovsky2001synchronization,njougouo2022synchronization}, have significantly contributed to advancing knowledge in various areas with the aim of enhancing human well-being. However, despite this progress, the complexity and diversity of natural systems continue to raise unresolved questions.

Among the cited collective behaviors, synchronization, introduced  by Huygens in 1673, is one of the most extensively studied phenomena in the literature due to its large applications in various real-life domains. For instance, in music, synchronization plays a crucial role in harmonizing vocal strings to produce pleasant sounds \cite{valdez2017revolver}. In medicine, it is used to coordinate the rhythm of a pacemaker's beats \cite{hartcher2016single} or in the study of brain diseases such as epilepsy \cite{missonnier2006decreased,gonccalves2015fresh}. In biological systems, for instance, collective behaviors are observed in the movements flocking birds \cite{ling2019costs, reynolds2022stochastic} and fish schooling \cite{pavlov2000patterns}. Similar collective behaviors are also seen in engineering applications, such as the coordinated motion control of drone swarms \cite{abdessameud2013motion, li2022truck}. In the same context, a particularly fascinating emerging concept is that of swarmalators--systems that combine swarming (spatial movement) and synchronization (temporal coordination) of interacting agents, as introduced by O’Keeffe et al. \cite{o2017oscillators}. These systems exist at the crossroads of synchronization and clustering \cite{o2017oscillators, sar2022dynamics, kongni2023phase, ghosh2024amplitude}.

However, although its applications are diverse and intriguing, its attainment depends on several factors, including the nature of the interactions between the entities or systems that form the network. These interactions can occur between two or more systems, and the position of these agents can be static or vary over time. In the latter case, the definition of how the interactions take place becomes crucial and thus gives the network a topology that varies over time. Many studies in network science focus on synchronization in networks with a fixed or static topology, i.e., the network structure remains constant over time \cite{belykh2008cluster,pikovsky2001synchronization,njougouo2022synchronization}. However, this assumption does not always accurately describe most systems, such as population dynamics, the movement of insects and social animals, and even humans. In reality, most living systems do not have a stationary topology, leading to questions about the impact of a non-stationary topology on achieving synchronization \cite{skinner2022topological,nagpal2008epithelial,ghosh2022synchronized,rakshit2017time}. In \cite{holme2012temporal,fujiwara2011synchronization}, the authors have proposed models with non-stationary topology and demonstrated the possibility of achieving synchronization. These configurations, where the structure of the network changes over time, were initially referred as moving neighborhood networks \cite{stilwell2006sufficient}. They exhibit a non-trivial and non-static interaction structure, implying that not all nodes are interconnected simultaneously, and the adjacency matrix evolves over time.

Mobile oscillators represent one of the most suitable applications of adaptive networks. In this scenario, each node comprises an oscillator associated with an agent moving in a space, with the coupling between the oscillators established based on the proximity of the agents' positions \cite{majhi2019emergence,malar2021multi,poklonskiy2022superradiation,xu2021collective,chowdhury2019synchronization}. Given the numerous applications of evolving networks, particularly in coordinating the movements of robots \cite{sun2016velocity}, vehicles \cite{cavraro2015data}, animals \cite{blonder2011time}, and even people \cite{blonder2011time}, this subject remains relevant. Exploring it would help us better understand or explain other behaviors observed in nature.

In 2017, G. Petrungaro et al. demonstrated that synchronization in mobile oscillators can be achieved by incorporating delayed coupling between oscillators and concluded that synchronization is enhanced at a lower mobility rate compared to instantaneous coupling \cite{petrungaro2019synchronization}. The transition to this synchronization in complex systems is a key phenomenon studied across various fields, including physics, biology, and engineering. Two types of transitions have been identified in mobile oscillators: the first-order (also refered to as explosive) transition  \cite{ling2020explosive,xiao2022explosive}, characterized by an abrupt and discontinuous change, and the second-order transition \cite{levis2017synchronization,perez2017control}, which is more gradual and continuous. 
This synchronization in mobile oscillators is significantly affected by the density of agents \cite{frasca2008synchronization} (i.e., number of agents per unit area), the timescale associated with changes in topology \cite{fujiwara2011synchronization} (When this timescale is longer than the one related to local synchronization within clusters, the system requires more time to achieve synchronization than predicted by the fast-switching approximation). Based on this, a relationship was established between the behavior of the systems and the properties of the dynamical network. The time scales between local dynamics and topology also change and have a significant influence on the global behavior of the systems. A more recent model of mobile systems has been proposed by Nguefoue et al. \cite{nguefoue2021network}, where both the agents' and oscillators' vision ranges are considered. Based on the influence of the internal dynamics on the position, in this model, the two parts of the system mutually influence each other, and the connectivity in one set strongly depends on the other. We observe that phase synchronization emerges through the formation of clusters in both the internal and external dynamics, describing a second-order transition.

In numerous natural and artificial systems, including social \cite{szell2010multirelational}, biological \cite{adhikari2011time}, and transportation \cite{cardillo2013modeling} systems, entities engage in complex interactions that often involve various types of relationships. Within these systems, a node within one network at a specific moment may concurrently belong to another network  \cite{sar2023flocking}. Such systems comprise multiple subsystems and layers of connectivity, making a multi-layer or multi-cluster architecture an effective framework for describing their complexity. Focusing on the animal world, mobile systems are employed to study the movement patterns of birds, wild animals, and fish, which often form flocks or schools for safety reasons \cite{sar2023flocking}. However, it is not uncommon to observe the formation of two or more groups of these animals, circling each other in phase synchrony, with individuals within each group exhibiting nearly identical behavior. Thus, while the clusters and the animals' bodies (represented by agents) are in spatial phase synchrony), their internal dynamics are fully synchronized. A particularly intriguing application involves studying flocks of birds in motion, where interactions among individuals shaped by their field of vision and ability to navigate through space give rise to complex collective behaviors such as alignment, cohesion, and the emergence of specific formations\cite{arrow2000small}. The introduction of a mobile multi-layer or multi-cluster framework allows for the integration of various interaction levels, such as social connections, spatial constraints, and responses to environmental disturbances, offering a more comprehensive understanding of the global dynamics and the underlying mechanisms driving these collective phenomena. An example could be the wildebeests - zebras interaction, where each species acts as a cluster of mobile elements.  These two species often coexist in mixed herds and can be viewed as "clusters" within a complex system, each with its own dynamics and behaviors, while interacting in subtle yet crucial ways for the group's survival. For instance, the quicker, more alert zebras are able to detect predators, while the less agile wildebeest rely on the zebras' vigilance for their safety, and in turn, provide an acute sense of smell that detect water sources. This interaction between the groups illustrates the concept of a coupled, multi-cluster system, where each cluster has distinct dynamics but also influences and interacts with the others\cite{huan,thaker}.

In this manuscript, we investigate two groups of non-interchangeable elements moving within each cluster, while both clusters are in motion in space. In this configuration, we analyze the collective behaviors of the nodes in both networks as a function of cluster velocity and inter-cluster coupling parameters. We then conduct a comparative study of the results obtained using the mobile systems model commonly found in the literature \cite{majhi2019emergence} and those obtained using the new model proposed by Nguefoue et al. \cite{nguefoue2021network}.

The remainder of the paper is structured as follows: Section \ref{sec2} introduces the model, which comprises two mobile clusters. In Section \ref{sec3}, we present the numerical results. We begin by examining the dynamics of the different clusters for various coupling parameters (Section \ref{sec3a}). Subsequently, we explore the mutual influence of the internal and external dynamics on the network's behavior. Finally, Section \ref{sec4} offers the conclusion of our work.

\section{Model}\label{sec2}
\begin{figure}[htp]
  \begin{center}
    \includegraphics[scale=0.3]{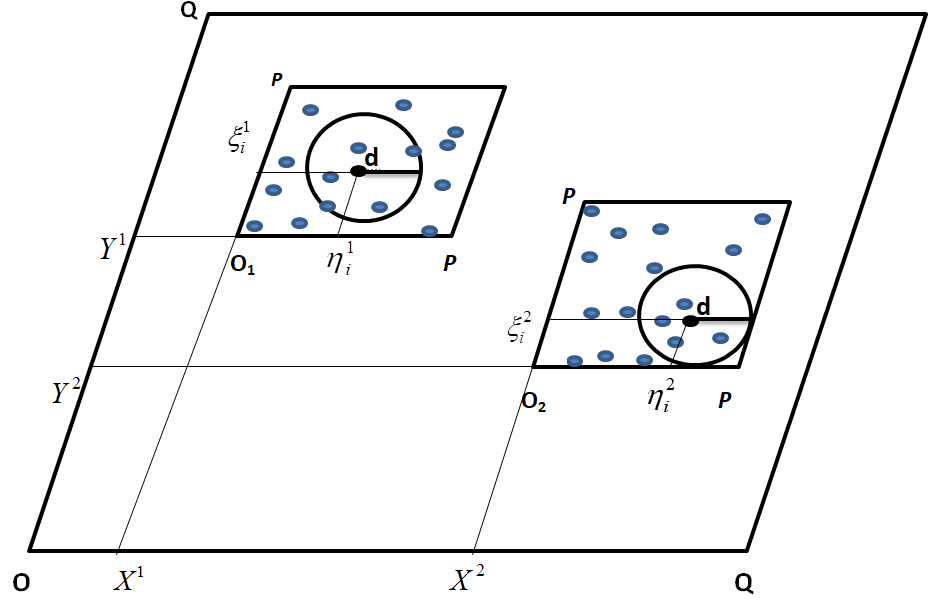}
    \caption{Relative positions of mobile agents in a two-dimensional space at a given time. The coordinates of each cluster are identified by the pair of coordinates $ \left( {X_i^k ,\,Y_i^k } \right)$  in space defined by $(Q,O,Q)$ and the individual's coordinates in the clusters are given by $\left( {\eta _i^k ,\,\xi _i^k } \right)$.}\label{fig1a}
  \end{center}
\end{figure}

The dynamic model presented in this study draws inspiration from the movement of animal groups that form sub-groups within a shared space, with each sub-group represented as a distinct cluster. The model is a network consisting of two two dimensional regions or clusters of mobile systems of dimensions $P \times P$, where they themselves are randomly moving relative to one another in a 2D space of dimensions $Q \times Q$. Each entity or agent, represented by blue dot points in each cluster (see Fig. \ref{fig1a}), represents a mobile system, also called an agent, moving randomly inside the region defined by the cluster. Each cluster, named $L_1$ and $L_2$ for the first and second region respectively, with periodic boundary conditions in the space $P \times P$, where the oscillator was generated, contains $N$ mobile systems. The 2D bounded space that includes the two 2D bounded sub-spaces $L_1$ and $L_2$ (i.e., $L = \{L_1, L_2\}$ of dimensions $P \times P$, such that $P \leq Q$ ), 
is also defined with periodic boundary conditions, and the motion (i.e., position) of clusters in this space is defined by Eq.\ref{eq32}.
\begin{equation}\label{eq32}
  \left\{ \begin{array}{l}
  {X^k}\left( {t + \Delta t} \right) = {X^k} + {v^k}\Delta t\cos \left( {{\varphi ^k}} \right),\,\,\bmod \left( Q-P \right)\\
  {Y^k}\left( {t + \Delta t} \right) = {Y^k} + {v^k}\Delta t\sin \left( {{\varphi ^k}} \right),\,\,\bmod \left( Q-P \right)
  \end{array} \right.
\end{equation}
The position of the $k^{th}$-cluster (with $k=1,2$) is defined by the pair of coordinates ($X^k, Y^k$), which defines the position of the lower-left corner of the $k^{th}$-cluster. At each time step $\Delta t$, the angle $\varphi^k$ takes values ranging from $-\frac{\pi}{2}$ to $\frac{\pi}{2}$, and $v^k$ represents the constant velocity of the $k^{th}$ cluster.

For simplicity, we assume that both clusters $L_1$ and $L_2$ have identical dimensions of $P\times P$. Let's define the position of each $i^{th}$ agent in a given $k^{th}$ cluster  by the pair of coordinates $\left( {\eta_i^k ,\xi_i^k } \right)$ as expressed in Eq.\ref{eq1}.
\begin{equation}\label{eq1}
  \left\{ \begin{array}{l}
  \eta _i^k\left( {t + \Delta t} \right) = \eta _i^k\left( t \right) + u_i^k\Delta t\cos \theta _i^k\left( t \right),\,\,\bmod \left( P \right)\\
  \xi _i^k\left( {t + \Delta t} \right) = \xi _i^k\left( t \right) + u_i^k\Delta t\sin \theta _i^k\left( t \right),\,\,\bmod \left( P \right)
  \end{array} \right.
\end{equation}
In each cluster of dimension $P\times P$, agents move randomly with a constant velocity $u_i^k$ ($i=1,2, ..., N$), and the direction of motion of each agent is determined by the angle $\theta_i^k(t)$. This angle is randomly chosen after every time step $\Delta t$ within a range of $-\frac{\pi}{2}$ to $\frac{\pi}{2}$. Consequently, the position of the $i^{th}$ agent in the $k^{th}$ region  at time $t$ is represented in the $2D$ space of dimension $Q \times Q$ by the coordinates $\left( {X_i^k ,\,Y_i^k } \right)$, with the corresponding mathematical expression given by Eq.\ref{eq2b}.
\begin{equation}\label{eq2b}
  \left\{ \begin{array}{l}
X_i^k\left( {t + \Delta t} \right) = \bmod \left( {{X^k}\left( {t + \Delta t} \right),\,Q - P} \right) \\ \,\,\,\,\,\,\,\,\,\,\,\,\,\,\,\,\,\,\,\,\,\,\,\,\,\,\,\,\,\,\,\,\,\,\,\,\,\,\,\,\,\,\,\,\,+ \bmod \left( {\eta _i^k\left( {t + \Delta t} \right),\, P} \right)\\
Y_i^k\left( {t + \Delta t} \right) = \bmod \left( {{Y^k}\left( {t + \Delta t} \right),\,Q - P} \right) \\
\,\,\,\,\,\,\,\,\,\,\,\,\,\,\,\,\,\,\,\,\,\,\,\,\,\,\,\,\,\,\,\,\,\,\,\,\,\,\,\,\,\,\,\,\,+ \bmod \left( {\xi _i^k\left( {t + \Delta t} \right),\,\, P} \right)
\end{array} \right.
\end{equation}

Let's define the internal dynamics of each agent in each cluster. This internal dynamics of the mobile system is described by chaotic oscillators, with the R\"ossler chaotic oscillator chosen for this study \cite{rossler1979continuous}. Thus, each agent is characterized by its internal dynamics (described by the R\"ossler oscillator) and its external dynamics (i.e., its position in the two-dimensional space). We define by ${\left( {{x^{1}_i}\left( t \right),\,{x^{2}_i}\left( t \right),\,{x^{3}_i}\left( t \right)} \right)}$ and $ {\left( {{y^{1}_i}\left( t \right),\,{y^{2}_i}\left( t \right),\,{y^{3}_i}\left( t \right)} \right)} $ the state variables of the $i^{th}$ oscillator in the first and second cluster respectively.
The spatial movement of agents induces interactions among them, driven by their proximity. In this study, these spatial or geographical proximities give rise to interactions within the internal dynamics. Therefore, the internal dynamics of the first  and second cluster can be described by Eq.\ref{eq3} and \ref{eq3s} respectively.
\begin{equation}\label{eq3}
\left\{ \begin{array}{l}
\dot x_i^1 =  - x_i^2 - x_i^3\\
\dot x_i^2 = x_i^1 + ax_i^2 + {\epsilon }\sum\limits_{j = 1}^N {g_{ij}^1(t)\left( {x_j^2 - x_i^2} \right)} \\
\,\,\,\,\,\,\,\,\,\,\,\,\,\,\,\,\,\,\,\,\,\,\,\,\,\,\,\,\,\,+ \mu {D_{xy}}\left( {\overline {y}^2   - x_i^2} \right) \\
\dot x_i^3 = b + x_i^3\left( {x_i^1 - c} \right)
\end{array} \right.
\end{equation}
%
\begin{equation}\label{eq3s}
\left\{ \begin{array}{l}
\dot y_i^1 =  - y_i^2 - y_i^3\\
\dot y_i^2 = y_i^1 + ay_i^2 + {\epsilon }\sum\limits_{j = 1}^N {g_{ij}^2(t)\left( {y_j^2 - y_i^2} \right)} \\
\,\,\,\,\,\,\,\,\,\,\,\,\,\,\,\,\,\,\,\,\,\,\,\,\,\,\,\,\,\,+ \mu {D_{yx}}\left( {\overline {x}^2   - y_i^2} \right) \\
\dot y_i^3 = b + y_i^3\left( {y_i^1 - c} \right)
\end{array} \right.
\end{equation}
where \(\epsilon\) and \(\mu\) are the intra- and inter-cluster  coupling coefficient respectively. $\overline{x}^2 = \frac{1}{m_1}\sum_{j=1}^{m_1} g_{ij}^1x_j^2$ and $\overline{y}^2 = \frac{1}{m_2}\sum_{j=1}^{m_2} g_{ij}^2y_j^2$ are the local center of mass for the first and second cluster, respectively. \( m_1 \) and \( m_2 \) denote the number of nodes within the vision size in the first and second cluster, respectively, as seen by an oscillator external to the cluster. This type of coupling between clusters operates under the hypothesis that an element in one cluster aligns with the collective movement of individuals in the other cluster, rather than the orientation of any specific individual. We consider $a=0.2$, $b=0.2$, and $c=5.7$ for all these cluster, and we note that at these values of the parameters the systems operate in the chaotic regime \cite{rossler1979continuous}. In this information exchange process, we consider bidirectional interactions within and between the clusters. The interaction between two oscillators $i$ and $j$ within the same cluster, is defined at each instant $t$ by the matrix $g_{ij}^k(t)$, which equals one when the Euclidean distance $d_{ij}^k(t)$ between them remains less than or equal to the agent vision size control (AVSC) parameter $d_0^k$ and zero elsewhere, as described in Eq.\ref{eq5}. This AVSC parameter represents the maximum distance within which an agent can detect and interact with another agent, essentially determining the range of an agent's perceptual field.
\begin{equation}\label{eq5}
 g_{ij}^k  = \left\{ \begin{array}{l}
 1\,\,\,if\,\,d_{ij}^k  \le d_0^k  \\
 0\,\,\,otherwise, \\
 \end{array} \right.
\end{equation}
with ${d^{k}_{ij}}\left( t \right) = \sqrt {{{\left( {{\eta^{k} _i}\left( t \right) - {\eta^{k} _j}\left( t \right)} \right)}^2} + {{\left( {{\xi^{k} _i}\left( t \right) - {\xi^{k} _j}\left( t \right)} \right)}^2}}$.

We assume that the interaction between the clusters occurs when the distance between the center of masses (defined above) \cite{hannachi2023moire} of the agents is within the vision size of the oscillators $d_{0}^k$.
Let's define $D_{XY}$ and $D_{YX}$ as the connectivity functions representing connections from a node in the first cluster to a node in the second cluster and vice versa, respectively. These functions are defined by:
\begin{equation}\label{eq6}
  D_{XY} = D_{YX}  = \left\{ \begin{array}{l}
 1\,\,\,if\,\,s_{XY},s_{YX}  \le s_0  \\
 0\,\,\,otherwise \\
 \end{array} \right.
\end{equation}
$s_{XY}$ denotes the Euclidean distance between two agents belonging to different clusters and is given by Eq.\ref{eq6a}. Here, $s_0$ is the maximum distance threshold of the clusters beyond which connectivity between agents from different clusters cannot be established.
\begin{equation}\label{eq6a}
  {s_{XY}} = {s_{YX}} = \sqrt {{{\left( {{X^1}\left( t \right) - {X^2}\left( t \right)} \right)}^2} + {{\left( {{Y^1}\left( t \right) - {Y^2}\left( t \right)} \right)}^2}}
\end{equation}

Therefore, the connectivity among oscillators within a cluster is contingent upon the closeness of the agents within that particular cluster, while the inter-cluster connectivity of the oscillators depends on the position of the agents' cluster in space, thus promoting rapid spreading of information within groups.

\section{Numerical results}\label{sec3}
In this section dedicated to the presentation of the numerical results of the model presented in section \ref{sec2}, we consider 50 agents in each cluster ($N = 50$) and this analysis can be extended to larger networks, as illustrated in Appendix~\ref{appB}. We set the surface areas of the agents' displacement spaces identical to $P\times P = 50\times 50$, and the surface area of the movement space of the flat spaces, denoted $P$, to $Q \times Q = 500 \times 500$. For simplicity, we consider the intra-cluster coupling strength between oscillators to be identical. We also assume identical vision sizes in both clusters and that the clusters move with the same speed, as do the agents within the clusters ( $d_0^k=d_0$, $v^k=v$, $u^k=u$, $k=1,2$). We used the fourth-order Runge-Kutta method for numerical integration of the oscillator system over $10^5$ iterations with a step of $dt = 0.05$.

\subsection{Dynamics of different clusters}\label{sec3a}
In this subsection, we aim to highlight the effect of the connectivity threshold between agents during their movements in the bounded space, as well as the coupling strength required to drive each cluster of the system towards a state of collective behavior, such as synchronization, which can be either phase or complete. Phase synchronization is characterized by the order parameter $r$, proposed by Kuramoto and Battogtokh \cite{kuramoto1991collective} where $r \to 1$ indicates synchronization and smaller values reflect desynchronization. This order parameter has also been detailed in \cite{njougouo2020dynamics, meli2023mobile} in the context of chaotic systems. The stability of complete synchronization is analyzed using the Master Stability Function (MSF) developed by Pecora and Carroll in 1998 \cite{pecora1998master}. Mathematical details for various systems can be found in \cite{pecora1998master, meli2023mobile}. In this work, we apply the MSF approach as detailed in the Appendix of ref.\cite{meli2023mobile} by the same authors, using the same system (Rössler) and configuration for intra-cluster dynamics. By applying this method to individual clusters, we are able to evaluate the Largest Lyapunov Exponent (LLE), which determines the stability of complete synchronization in the systems describing our internal dynamics (specifically, the Rössler system). Thus, if the LLE is greater than zero, synchronization is considered unstable, whereas if the LLE is less than zero, we can conclude that synchronization is stable.

\begin{figure}[htp]
  \begin{center}
    \includegraphics[scale=0.22]{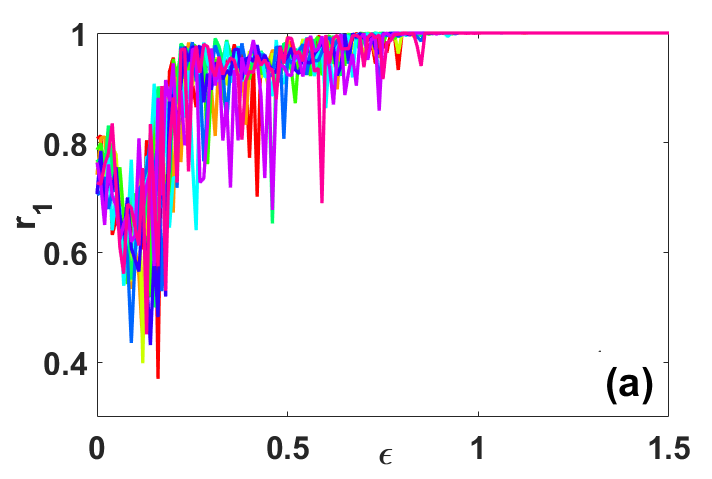}
    \includegraphics[scale=0.22]{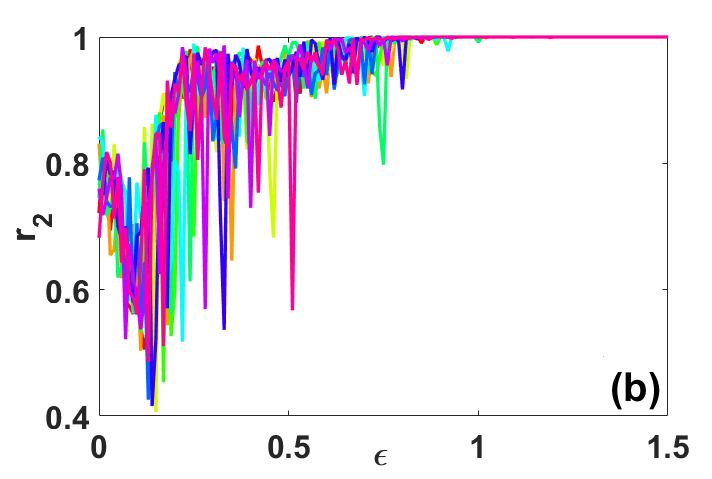}
    \includegraphics[scale=0.22]{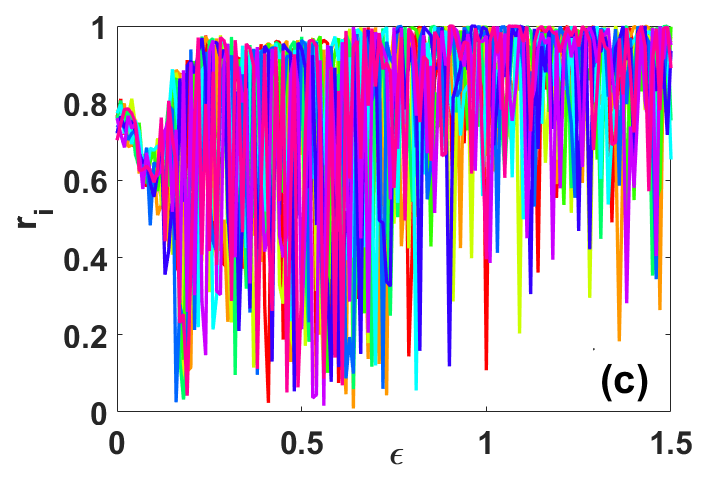}\\
    \includegraphics[scale=0.22]{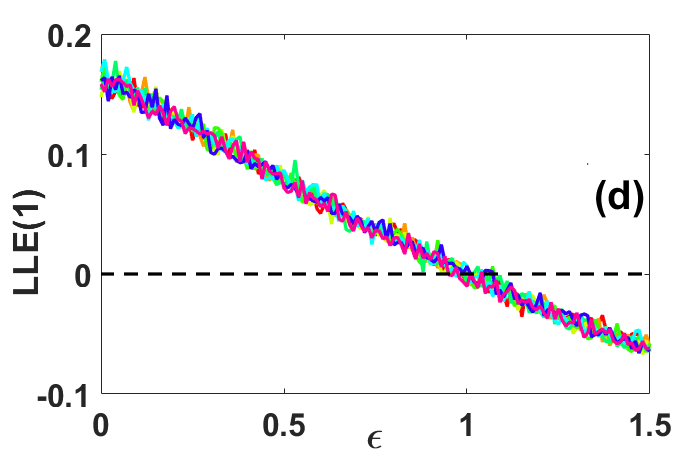}
    \includegraphics[scale=0.22]{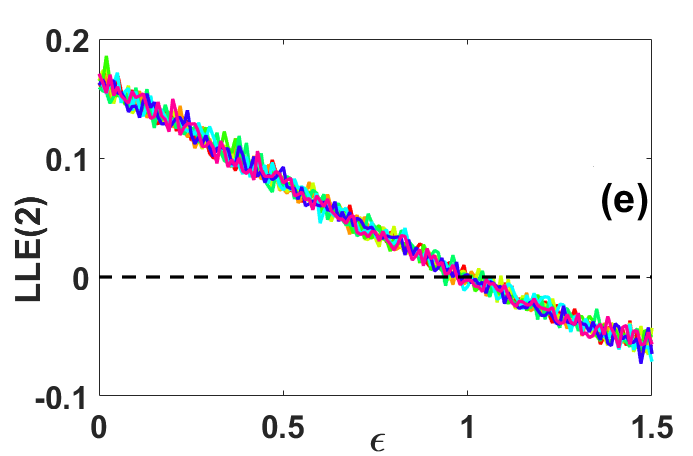}
    \caption{\footnotesize{Intra and inter-cluster dynamics of the internal state of the network as a function of the intra-cluster coupling strength $\epsilon$ for ten (10) different initial conditions in both internal and external parts of the system. (a) the first cluster's order parameter; (b), the second cluster's; (c) inter-cluster order parameter. (d) evolution of  LLE of the first cluster;  (e) ibid for the second cluster} as a function of $\epsilon$. Other parameters: $\mu = 1$, $u = 2$, $v = 10$, $d_{0} = 2$ .
    }\label{fig01}
  \end{center}
\end{figure}

Let's illustrate in Fig.\ref{fig01} the collective dynamics within (see Fig.~\ref{fig01}(a,b,d,e)) and between the clusters (see Fig.~\ref{fig01}(c)) as a function of the intra-clusters coupling strength $\epsilon$. For ten different initial conditions for both the positions and the oscillators, we present the order parameter characterizing phase synchronization in each cluster (see Fig.~\ref{fig01}(a) and Fig.~\ref{fig01}(b) for the first and second cluster, respectively) and between the clusters (see Fig.~\ref{fig01}(c)). Similarly, complete synchronization in the first and second clusters is respectively depicted in Fig.~\ref{fig01}(d) and Fig.~\ref{fig01}(e). The different colors observed correspond to each initial condition used.
Despite the random process in the connection between agents (i.e., variation of the adjacency matrices over time), it is interesting to observe that for all ten initial conditions, the intra-cluster dynamics remain relatively unchanged, as well as for the inter-cluster dynamics.

\begin{figure}[htp]
  \begin{center}
    \includegraphics[scale=0.22]{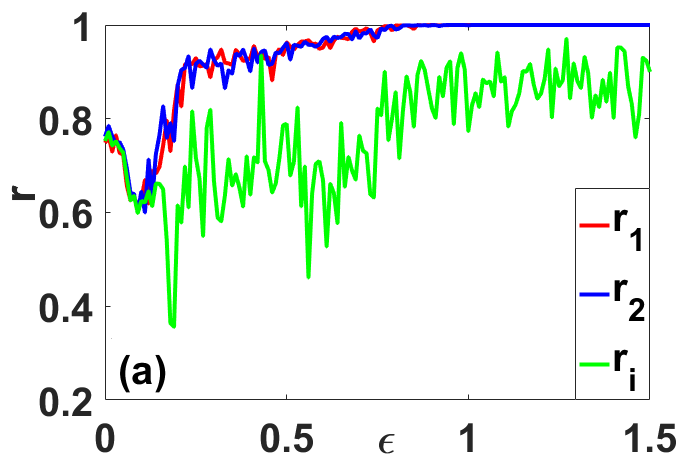}
    \includegraphics[scale=0.22]{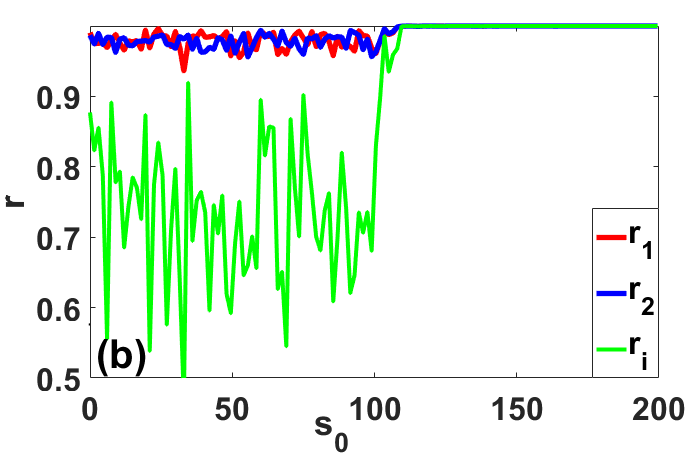}
    \includegraphics[scale=0.22]{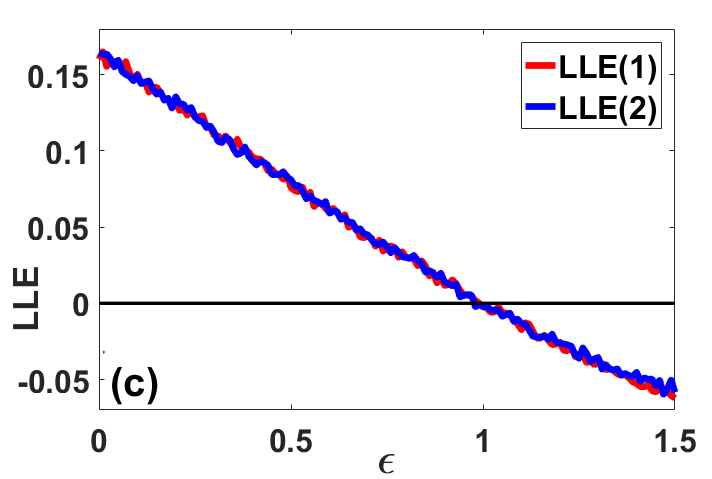}
    \includegraphics[scale=0.22]{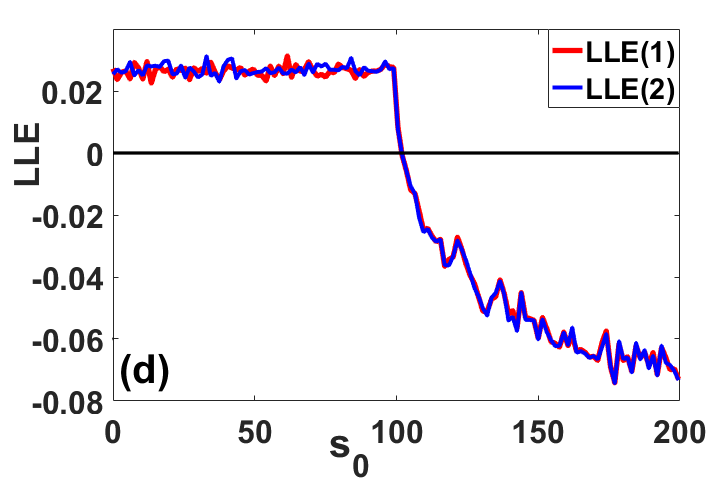}
    \caption{\footnotesize{The influence of intra-cluster coupling strength ($\epsilon$) and connectivity threshold between both clusters ($s_0$) on the network dynamics is investigated. By varying $\epsilon$, we present in (a) the order parameter depicting the collective dynamics of the phase of oscillators, and in (c) the Largest Lyapunov Exponent (LLE) showing the collective dynamics in terms of complete synchronization both within the same range of $\epsilon$ with $s_0=100$. Similarly, in (b) and (d), we analyze $s_0$ as the control parameter with $\epsilon = 0.8$. The red curve represents the dynamics of the first cluster, the blue curve represents the second cluster's dynamics, and the green curve illustrates the inter-cluster dynamics. Other parameter: $\mu = 1$, $u = 2$, $v = 10$, $d_{0} = 2$ .}
    }\label{fig1}
  \end{center}
\end{figure}

\begin{figure}[htp]
\centering
\begin{tabular}{cc}
\includegraphics[width=0.25\textwidth]{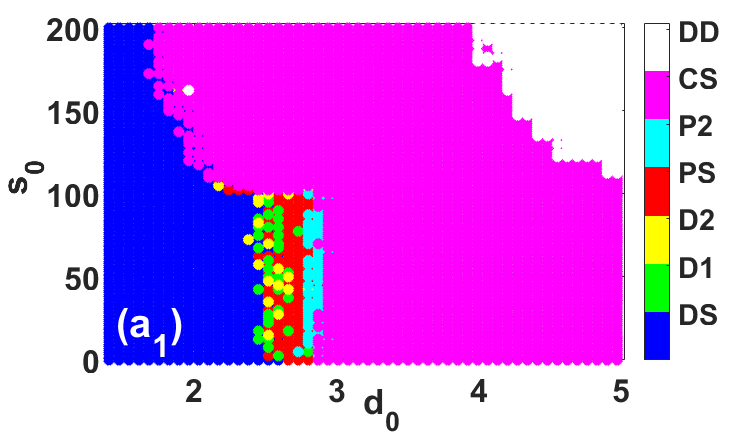}&
\includegraphics[width=0.25\textwidth]{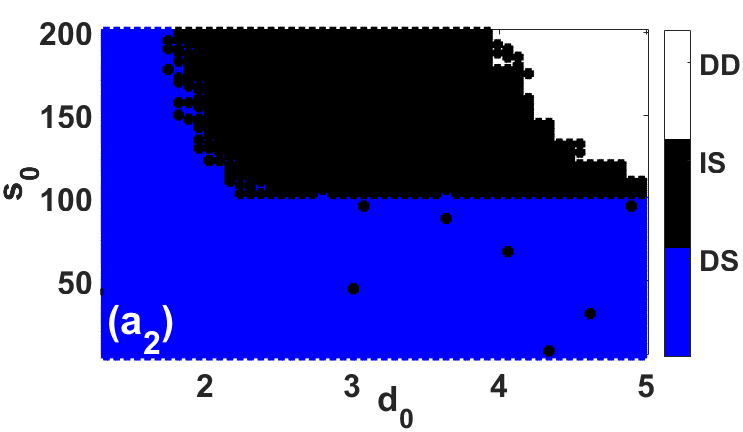}\\
\includegraphics[width=0.25\textwidth]{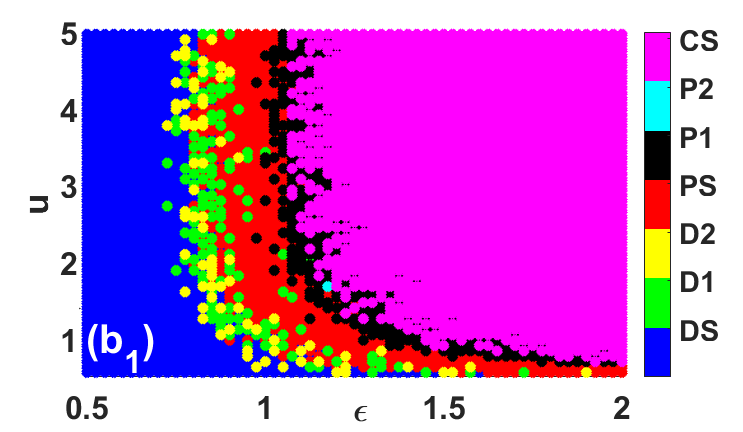}&
\includegraphics[width=0.25\textwidth]{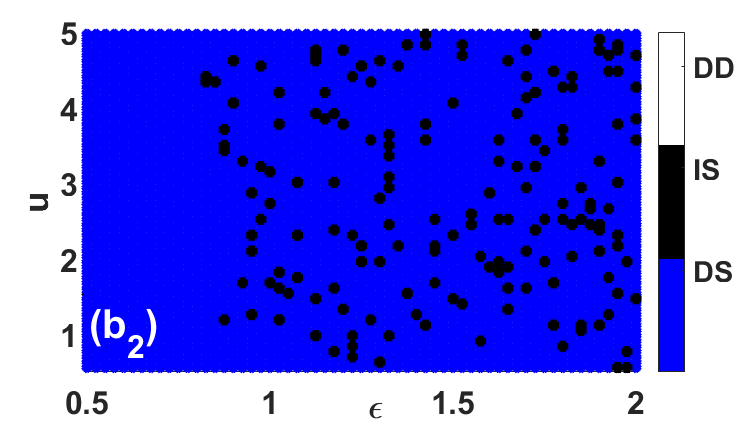}\\
\includegraphics[width=0.25\textwidth]{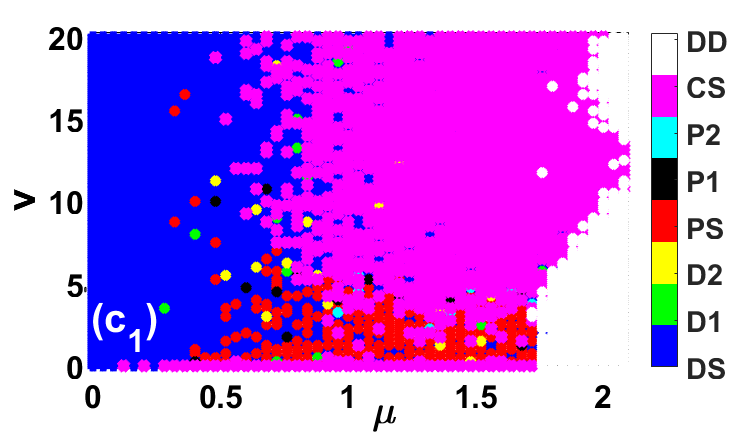}&
\includegraphics[width=0.25\textwidth]{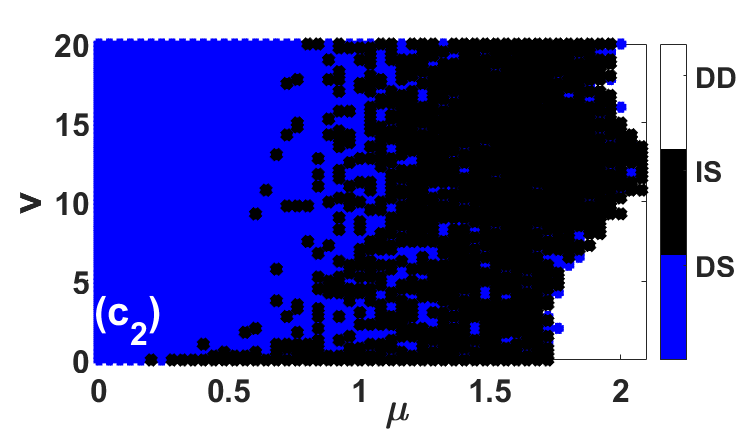}\\
\end{tabular}
\caption{\footnotesize{Different collective states achieved by the internal state (i.e., oscillators) of the network, which depend on the following parameters: ($a_1$, $a_2$) ${d_0}$ and ${s_0}$ with $\mu =1$, $\epsilon =0.5$, $v=5$ and $u=2$; ($b_1$, $b_2$) $\epsilon $ and $u$ with $\mu =1$, ${s_0}=100$, ${d_0}=2$ and $v=2$; ($c_1$, $c_2$) $\mu$ and $v$ ($\epsilon =0.5$, ${d_0}=2$, ${s_0}=100$ and $u=2$). The meanings of the different colors and acronyms are provided in Table.\ref{tab1}.}  
}
\label{fig3}
\end{figure}

For a more streamlined presentation of these findings, we present in Fig.~\ref{fig1} the averaged values of $r$ and LLE computed over the ten simulations, each representing distinct initial conditions. These averages are plotted against incremental changes in the control parameters: $\epsilon$ (depicted in Fig.~\ref{fig1}(a) and (c) for intra-cluster coupling strength) and $s_0$ (illustrated in Fig.~\ref{fig1}(b) and (d) for the threshold distance of connectivity between the two clusters). In this figure, the colors red, blue, and green respectively denote the dynamics of the first cluster, second cluster, and inter-cluster dynamics.
By fixing the other parameters of the system such as: $\mu = 1$, $u = 2$, $v = 10$, $d_{0} = 2$, and $s_0=100$, an increase in the coupling strength between the agents gradually leads the systems in the different clusters towards phase synchronization and even complete synchronization, as illustrated in Fig.\ref{fig1}(a,c). However, despite this increase, achieving synchronization, whether phase or complete, between the two clusters of the network proves to be very difficult. This result seems legitimate because to easily drive the entire system towards a state of synchronization, the appropriate control parameter remains either the inter-cluster coupling strength or the threshold distance of connectivity between the two clusters (see Fig.\ref{fig1}(b,d) with $\epsilon=0.8$). Therefore, increasing the threshold distance means increasing the probability of connecting more systems, thus raising the chances of synchronizing them. However, without any interaction (connectivity), achieving synchronization is a pure chance.\\
Let's explore the impact of the following combined parameters on the collective behavior of the entire system: the mobility parameters of the clusters $\left(s_0, v\right)$, the mobility parameters of the agents $\left(d_0, u\right)$, and the coupling parameters within and between clusters, denoted as $\epsilon$ and $\mu$ respectively. The intra and inter-cluster dynamics, combining these various parameters, are depicted in Fig.\ref{fig3}. Different behaviors are represented by different colors, the meanings of which are specified in Table.\ref{tab1}.

\begin{table}[h]
\centering
\caption{Characterized states and their representations}
\label{tab1}
\begin{tabular}{|p{3cm}|p{5cm}|}
\hline
Representations & Characterized states \\ \hline
DS (blue) & Disorder state in both clusters\\ \hline
D1 (green) & Disorder state only in the first cluster\\ \hline
D2 (yellow) & Disorder state only in the second cluster\\ \hline
PS (red) & Phase synchronization in both clusters\\ \hline
P1 (black, first column) & Phase synchronization only in the first cluster \\ \hline
P2 (cyan) & Phase synchronization only in the second cluster\\ \hline
CS (magenta) & Complete synchronization in both clusters\\ \hline
IS (black, second column) & Phase synchronization between the clusters\\ \hline
DD (white) & Divergence state \\ \hline
\end{tabular}
\end{table}

The first column of Fig.\ref{fig3} (i.e., Fig.\ref{fig3}($a_1, b_1, c_1$)) presents the different collective behaviors observed in the internal dynamics (i.e., oscillators) of the different clusters that form the network. These figures globally depict, based on various parameters, the different states through which the system passes from disorder in both clusters (DS) to complete synchronization in both clusters (CS). Examining the connectivity thresholds of agents $d_0$ and clusters $s_0$ (see Fig.\ref{fig3}($a_1$)), while keeping the other parameters constant ($\mu = 1$, $\epsilon = 0.5$, $v = 5$, and $u^{k} = 2$), it is observed that when $s_0 \leq 100$, the transition from the DS state to the CS state often passes through multiple states, namely: $D_1$, $D_2$, PS, and $P_2$, in ascending order of $d_0$. However, as $s_0$ becomes larger and larger (i.e., $s_0 > 100$), we observe a direct transition from the DS state to the CS state. This result, in general, once again underscores the previously drawn conclusion: the larger the threshold becomes, the more systems interact (i.e., are connected), thereby greatly increasing the likelihood of rapid synchronization. Regarding the inter-cluster dynamics (see the second column, i.e., Fig.\ref{fig3}($a_2, b_2, c_2$)), we primarily observe three behaviors: disorder (DS), synchronization between clusters (IS), and divergence state (DD), which corresponds to a lack of system solutions or system breakdown. Hence, it becomes apparent that achieving phase synchronization between the two clusters is facilitated when the opportunity for exchange between the two clusters greatly increases (i.e., $s_0 > 100$, see Fig.\ref{fig3}($a_2$)).

For fixed values of intra and inter-clusters connectivity thresholds, with ${d_0}=2$ and $s_0 =100$, and considering other parameters as $\mu =1$ and $u^{k}=2$, the system as a whole is expected to evolve towards an asynchronous state in the absence of $\epsilon$ and $u$. This intra-clusters desynchronization state arises from a low probability (low chance of interaction) for systems to interact and synchronize their internal dynamics. The results from Fig.\ref{fig3}($b_1$), despite variations in parameters specific to dynamics within different clusters, such as the agents' movement speed $u$ and the coupling strength between oscillators within the same cluster $\epsilon$, demonstrate that with significant effort (i.e., increasing $\epsilon$) or much greater mobility (i.e., increasing $u$), a state of complete synchronization can be achieved within both clusters. However, this intra-cluster synchronization does not guarantee synchronization between clusters, as illustrated in Fig.\ref{fig3}($b_2$), where the black color (IS), representing the phase synchronization zone between clusters, does not follow any coherent process when varying both control parameters.

Fig.\ref{fig3}($c_1$) presents the effect of inter-cluster coupling strength $\mu$ and the velocity of movement of different clusters $v$ on the dynamics within the clusters, with other parameters fixed at $\epsilon =0.5$, ${d_0}=2$, ${s_0}=100$, and $u^{k}=2$. Depending on the value of velocity of the clusters (i.e., $v$), the transition from the state of disorder (desynchronization) to synchronization in both clusters (CS) can occur directly (from desynchronization to synchronization directly), or it may pass through other behaviors such as $D_1$, $D_2$, $PS$, $P_1$, and $P_2$, although some behaviors are almost nonexistent. However, these inter-cluster parameters promote synchronization between clusters, as shown in Figure \ref{fig3}($c_2$).

\subsection{Stability of synchronization of the oscillators}\label{sec3b}

Let's analyze the stability of the synchronization of the internal dynamics (i.e., oscillators) of the whole system using the Lyapunov function \citep{kakmeni2010practical,sekieta1996practical}. The goal is to determine the conditions under which the synchronization of the internal part of the whole system is achieved in order to compare analytical results with numerical findings. We define the error system using the formula $e_i^k = x_i^k - y_i^k$ with $k=1,2,3$ and $i=1,2,...,N$, and its expression is given by Eq.\ref{st1}.

\begin{equation}\label{st1}
  \left\{ \begin{array}{l}
\dot e_i^1 =  - e_i^2 - e_i^3\\
\dot e_i^2 = e_i^1 + ae_i^2 + \epsilon \sum\limits_j^N {{g_{ij}^1\left( {e_j^2 - e_i^2} \right) }} + \epsilon {f_{1i}} + \\
\,\,\,\,\,\,\,\,\,\,\,\,\,\,\,\mu {D_{xy}}\left[ { - e_i^2 - \frac{1}{{{m_2}}}\sum\limits_j^{{m_2}} {g _{ij}^2e_j^2}  + {f_{2i}}} \right]\\
\dot e_i^3 =  - ce_i^3 + e_i^1y_i^3 + e_i^3y_i^1
\end{array}, \right.
\end{equation}
where ${f_{1i}} = \sum\limits_j^{{N}} {\left( {g_{ij}^1 - g_{ij}^2} \right)\left( {y_j^2 - y_i^2} \right)}$ and
${f_{2i}} = \frac{1}{{{m_1}}}\sum\limits_j^{{m_1}} {g _{ij}^1y_j^2}  - \frac{1}{{{m_2}}}\sum\limits_j^{{m_2}} {g _{ij}^2y_j^2}$. In this Eq.\ref{st1}, the non-linear term ${e^1_i}{e^3_i}$ of the third variable is disregarded due to its negligible size, approaching zero as the system approaches synchronization.\\

Let's consider the function $V_i$ defined by Eq.\ref{st2}, regarded as a Lyapunov function candidate.

\begin{equation}\label{st2}
  {V_i} = \frac{1}{2}\left( {{{\left( {e_i^1} \right)}^2} + {{\left( {e_i^2} \right)}^2} + {{\left( {e_i^3} \right)}^2}} \right) + \int\limits_0^t {\left( {{\gamma _1}{{\left( {e_i^1} \right)}^2} + {\gamma _2}{{\left( {e_i^3} \right)}^2}} \right)dt}
\end{equation}
Where ${\gamma _1}$ and ${\gamma _2}$ are parameters to be determined. The derivative of this Lyapunov function candidate with respect to time yields:

\begin{equation}\label{st4}
  \begin{array}{l}
{{\dot V}_i} = \left( {-1 + y_i^3} \right)e_i^1e_i^3 + a{\left( {e_i^2} \right)^2} - c{\left( {e_i^3} \right)^2} + y_i^1{\left( {e_i^3} \right)^2} + {\gamma _1}{\left( {e_i^1} \right)^2}\\
 - {\gamma _1}{\left( {e_i^1\left( 0 \right)} \right)^2} + {\gamma _2}{\left( {e_i^3} \right)^2} - {\gamma _2}{\left( {e_i^3\left( 0 \right)} \right)^2} + \epsilon \sum\limits_{j \ne i}^N {g_{ij}^1e_j^2e_i^2}  - \\
\epsilon {\left( {e_i^2} \right)^2}\sum\limits_{j \ne i}^N {g_{ij}^1}  + e_i^2\epsilon {{f_{1i}}}  - \mu {D_{xy}}\left( {1 + \frac{1}{{{m_2}}}} \right){\left( {e_i^2} \right)^2} - \\
\frac{{\mu {D_{xy}}}}{{2{m_2}}}\sum\limits_j^{{m_2}} {g _{ij}^2e_j^2e_i^2}  + e_i^2\mu {D_{xy}}{f_{2i}}
\end{array}
\end{equation}
Let's assume $y_m^1 = \max\left| {y_i^1} \right|$ and $y_m^3 = \max\left| {y_i^3} \right|$. Therefore, the derivative of the Lyapunov function is maximized as follows:

\begin{equation}\label{st5}
  \begin{array}{l}
{{\dot V}_i} \le \left( {1 + y_m^3} \right)\left| {e_i^1} \right|\left| {e_i^3} \right| + a{\left( {e_i^2} \right)^2} - c{\left( {e_i^3} \right)^2} + y_m^1{\left( {e_i^3} \right)^2} + \\
{\gamma _1}{\left( {e_i^1} \right)^2} - {\gamma _1}{\left( {e_i^1\left( 0 \right)} \right)^2} + {\gamma _2}{\left( {e_i^3} \right)^2} - {\gamma _2}{\left( {e_i^3\left( 0 \right)} \right)^2} + \\
\epsilon \sum\limits_{\scriptstyle j = 1,\hfill\atop \scriptstyle j \ne i\hfill}^N {g_{ij}^1\left| {e_j^2} \right|\left| {e_i^2} \right|}  - \epsilon {\left( {e_i^2} \right)^2}\sum\limits_{\scriptstyle j = 1,\hfill\atop \scriptstyle j \ne i\hfill}^N {g_{ij}^1}  - \mu {D_{xy}}\left( {1 + \frac{1}{{{m_2}}}} \right){\left( {e_i^2} \right)^2}\\
 + \frac{{\mu {D_{xy}}}}{{2{m_2}}}\sum\limits_j^{{m_2}} {g _{ij}^2\left| {e_j^2} \right|\left| {e_i^2} \right|} + \epsilon \left| {e_i^2} \right|\left| {{f_{1i}}} \right| + \mu {D_{xy}}\left| {e_i^2} \right|\left| {{f_{2i}}} \right|
\end{array}
\end{equation}
The transition from Eq.\ref{st5} to Eq.\ref{st6} considers the following mathematical property. $\left| a \right|\left| b \right| \le \frac{1}{2}\left( {{a^2} + {b^2}} \right)$, with $a$ and $b$ the real numbers.

\begin{equation}\label{st6}
  \begin{array}{l}
{{\dot V}_i} \le \frac{1}{2}\left( {1 + y_m^3} \right){\left( {e_i^1} \right)^2} + \,\left( {y_m^1 - c + \frac{1}{2}\left( {1 + y_m^3} \right)} \right){\left( {e_i^3} \right)^2} +\\
\left[ {a + \frac{\epsilon }{2} - \mu {D_{xy}}\left( {\frac{1}{2} + \frac{1}{{{m_2}}} - \frac{1}{{4{m_2}}}\sum\limits_{\scriptstyle j = 1,\hfill\atop \scriptstyle j \ne i\hfill}^{{m_2}} {g _{ij}^2} } \right) - \frac{\epsilon }{2}\sum\limits_{\scriptstyle j = 1,\hfill\atop \scriptstyle j \ne i\hfill}^N {g_{ij}^1} } \right]{\left( {e_i^2} \right)^2}\\
 + {\gamma _1}{\left( {e_i^1} \right)^2} - {\gamma _1}{\left( {e_i^1\left( 0 \right)} \right)^2} + {\gamma _2}{\left( {e_i^3} \right)^2} - {\gamma _2}{\left( {e_i^3\left( 0 \right)} \right)^2} + \frac{\epsilon }{2}{\left( {{f_{1i}}} \right)^2}\\
 + \frac{\epsilon }{2}\sum\limits_{\scriptstyle j = 1,\hfill\atop \scriptstyle j \ne i\hfill}^N {g_{ij}^1{{\left( {e_j^2} \right)}^2}}  + \frac{{\mu {D_{xy}}}}{{4{m_2}}}\sum\limits_{\scriptstyle j = 1,\hfill\atop \scriptstyle j \ne i\hfill}^{{m_2}} {g _{ij}^2{{\left( {e_j^2} \right)}^2}} + \frac{{\mu {D_{xy}}}}{2}{\left( {{f_{2i}}} \right)^2}
\end{array}
\end{equation}
As all $m_2$ oscillators within the visual size are inherently connected to oscillator $i$, the degree of this node is naturally $m_2$ and then, $\frac{1}{{4{m_2}}}\sum\limits_{\scriptstyle j = 1,\hfill\atop \scriptstyle j \ne i\hfill}^{{m_2}} {g _{ij}^2}  = \frac{1}{4}$.

Let us assume $Z_1$ and $Z_2$ be real positive, then $\,\,\sum\limits_{\scriptstyle j = 1,\hfill\atop \scriptstyle j \ne i\hfill}^N {g_{ij}^1{{\left( {e_i^2} \right)}^2}}  \le {Z_1}{\left( {e_i^2} \right)^2}$ and $\,\,\sum\limits_{\scriptstyle j = 1,\hfill\atop \scriptstyle j \ne i\hfill}^N {g _{ij}^2{{\left( {e_i^2} \right)}^2}}  \le {Z_2}{\left( {e_i^2} \right)^2}$.\\ 
If we assume that there exists a $Z$ such that $Z = \max\left( {Z_1, Z_2} \right)$, then, we have:
\begin{equation}\label{qqq}
  \left[ {\frac{\epsilon }{2}{Z_1} + \frac{{\mu {D_{xy}}}}{{2{m_2}}}{Z_2}} \right]{\left( {e_i^2} \right)^2} \le \left[ {\frac{\epsilon }{2} + \frac{{\mu {D_{xy}}}}{{2{m_2}}}} \right]Z{\left( {e_i^2} \right)^2}.
\end{equation}
\\
Assuming ${\gamma _1} = \frac{1}{2}\left( {1 + y_m^3} \right)$ and ${\gamma _2} = y_m^1 - c + \frac{1}{2}\left( {1 + y_m^3} \right)$, Eq.\ref{st6} is transformed into Eq.\ref{st8}.

\begin{equation}\label{st8}
  \begin{array}{l}
{{\dot V}_i} \le 2{\gamma _1}{\left( {e_i^1} \right)^2} - {\gamma _1}{\left( {e_i^1\left( 0 \right)} \right)^2} + 2{\gamma _2}{\left( {e_i^3} \right)^2} - {\gamma _2}{\left( {e_i^3\left( 0 \right)} \right)^2} +\\
\left[ {a + \frac{\epsilon }{2} + \frac{1}{2}\left( {\epsilon  + \frac{{\mu {D_{xy}}}}{{{m_2}}}} \right)Z - \frac{\mu {D_{xy}}}{{{m_2}}}  - \frac{\epsilon }{2}\sum\limits_{\scriptstyle j = 1,\hfill\atop \scriptstyle j \ne i\hfill}^N {g_{ij}^1} } \right]{\left( {e_i^2} \right)^2}\\
+ \frac{\epsilon }{2}{\left( {{f_{1i}}} \right)^2} + \frac{{\mu {D_{xy}}}}{2}{\left( {{f_{2i}}} \right)^2}
\end{array}
\end{equation}
Given ${\gamma}$ defined as $\gamma  = \max\left( {{\gamma _1},\,{\gamma _2}} \right)$, we can express $\dot{V}_i$ as:
\begin{equation}\label{st9}
  \begin{array}{l}
{{\dot V}_i} \le 2\gamma {\left\| {{e_i}} \right\|^2} - \gamma {\left\| {{e_i}\left( 0 \right)} \right\|^2} + \frac{\epsilon }{2}{\left( {{f_{1i}}} \right)^2} + \frac{{\mu {D_{xy}}}}{2}{\left( {{f_{2i}}} \right)^2} +\\
 \left[ {a + \frac{1}{2}\left( {\epsilon  + \frac{{\mu {D_{xy}}}}{{{m_2}}}} \right)Z - \mu {D_{xy}}\left( {\frac{1}{2} + \frac{1}{{{m_2}}}} \right) - \frac{\epsilon }{2}\sum\limits_{\scriptstyle j = 1,\hfill\atop \scriptstyle j \ne i\hfill}^N {g_{ij}^1} } \right]{\left( {e_i^2} \right)^2}
\end{array}
\end{equation}

Our Lyapunov function candidate, defined in Eq.\ref{st2}, can be considered as a Lyapunov function if $ \dot V_i \le 0 $ and $\dot V_i=0$ when ${e_i^k}=0$. Therefore, the stability conditions for synchronization between clusters are provided by Eq.\ref{st10}, derived from Eq.\ref{st9}.
\begin{equation}\label{st10}
  \begin{array}{l}
2\gamma {\left\| {{e_i\left( t \right)}} \right\|^2} - \gamma {\left\| {{e_i}\left( 0 \right)} \right\|^2} \le 0\\
a + \frac{1}{2}\left( {\epsilon  + \frac{{\mu {D_{xy}}}}{{{m_2}}}} \right)Z - \mu {D_{xy}}\left( {\frac{1}{2} + \frac{1}{{{m_2}}}} \right) - \frac{\epsilon }{2}\sum\limits_{\scriptstyle j = 1,\hfill\atop \scriptstyle j \ne i\hfill}^N {g_{ij}^1}  \le 0\\
{f_{1i}}=0\\
{f_{2i}}=0
\end{array} 
\end{equation}
From Eq.\ref{st10}, it follows that ${f_{1i}}=0$ and ${f_{2i}}=0$, implying $y_j^2 = y_i^2$ for all $i$ and $j$. In other words, the oscillators within a cluster synchronize, and the number of oscillators connected to the $i^{th}$ oscillator are equal. Let's consider the set of relations given by Eq.\ref{st10a}.

\begin{equation}\label{st10a}
 \begin{array}{l}
fe_{max}=max(2 {\left\| {{e_i\left( t \right)}} \right\|^2} -  {\left\| {{e_i}\left( 0 \right)} \right\|^2}) \\
f1_{max}=max(f_{1i}) \\
f2_{max}=max(f_{2i}) \\
f3_{max}=max(a + \frac{1}{2}\left( {\epsilon  + \frac{{\mu {D_{xy}}}}{{{m_2}}}} \right)Z - \mu {D_{xy}}\left( {\frac{1}{2} + \frac{1}{{{m_2}}}} \right) \\ 
\,\,\,\,\,\,\,\,\,\,\,\,\,\,\,\,\,\,\,\,\,\,\, - \frac{\epsilon }{2}\sum\limits_{\scriptstyle j = 1,\hfill\atop \scriptstyle j \ne i\hfill}^N {g_{ij}^1})
\end{array} .
\end{equation}
From this maximized system described in Eq.\ref{st10a}, we can summarize the conditions required to ensure the stability of synchronization as follows:
\begin{equation}\label{st10aaa}
 \begin{array}{l}
fe_{max} \leq 0\\
f1_{max} = 0 \\
f2_{max} = 0\\
f3_{max} \leq 0\
\end{array} .
\end{equation}

\begin{figure}[htp]
  \begin{center}
    \includegraphics[scale=0.2]{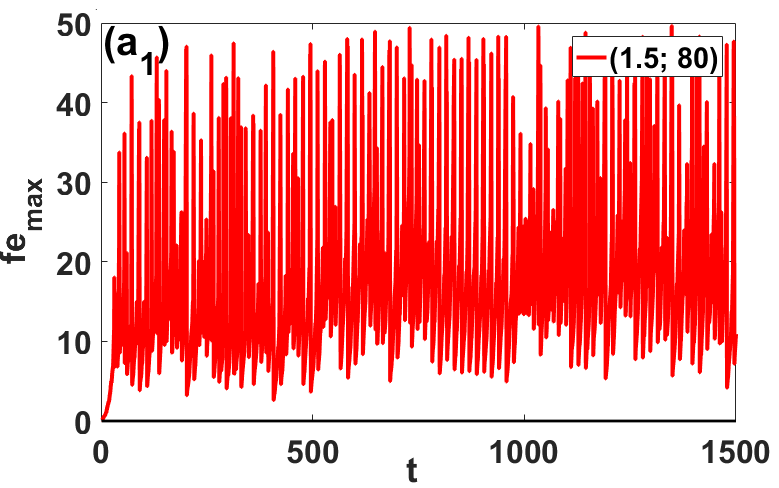}
    \includegraphics[scale=0.2]{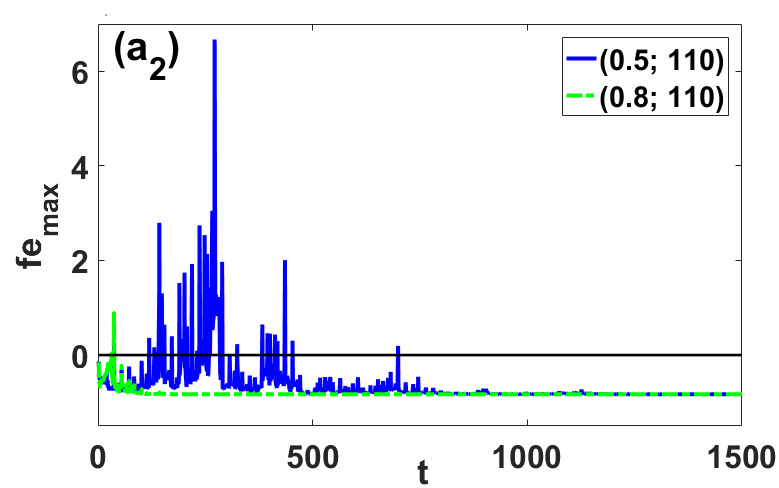}
    \includegraphics[scale=0.2]{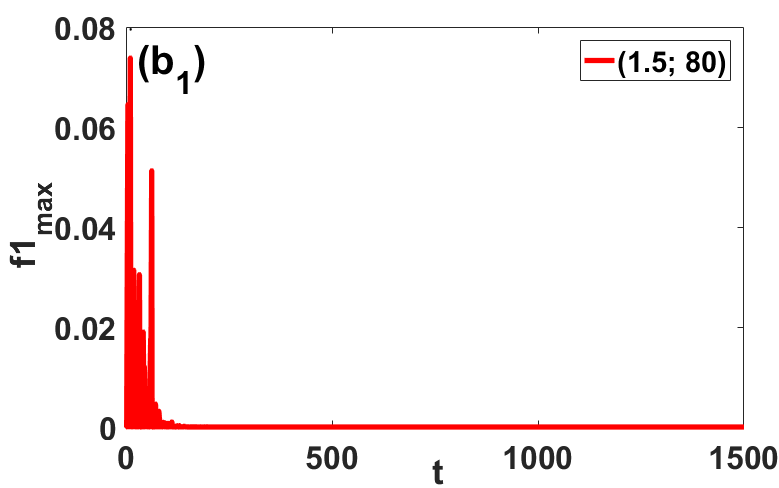}
    \includegraphics[scale=0.2]{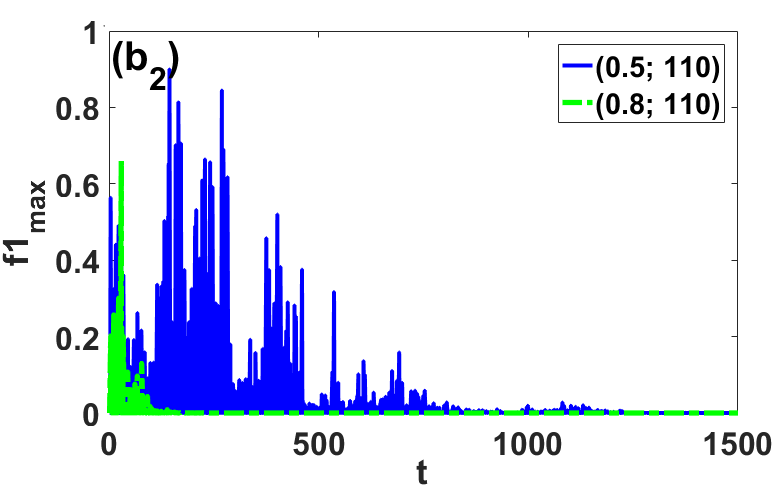}
    \includegraphics[scale=0.2]{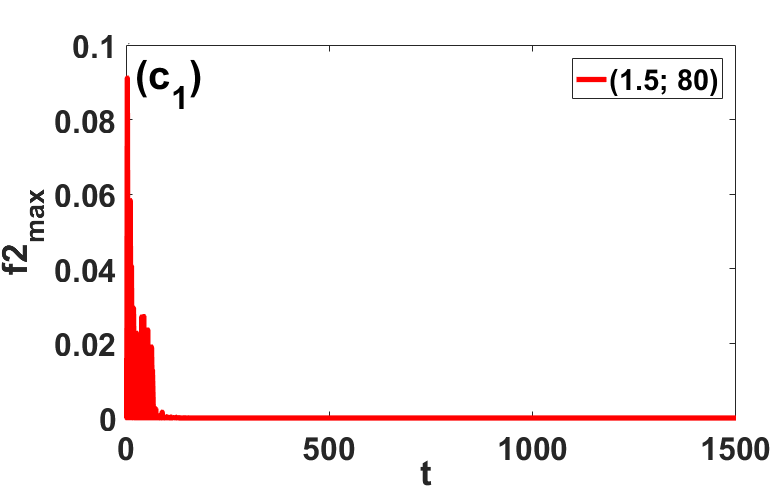}
    \includegraphics[scale=0.2]{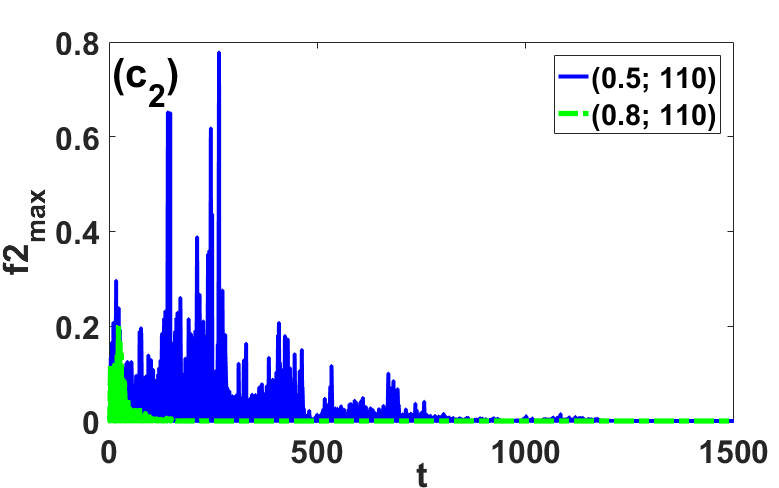}
    \includegraphics[scale=0.2]{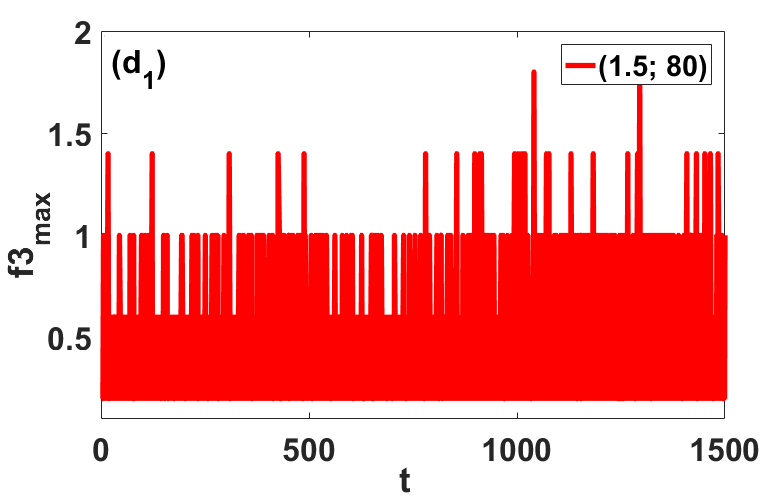}
    \includegraphics[scale=0.2]{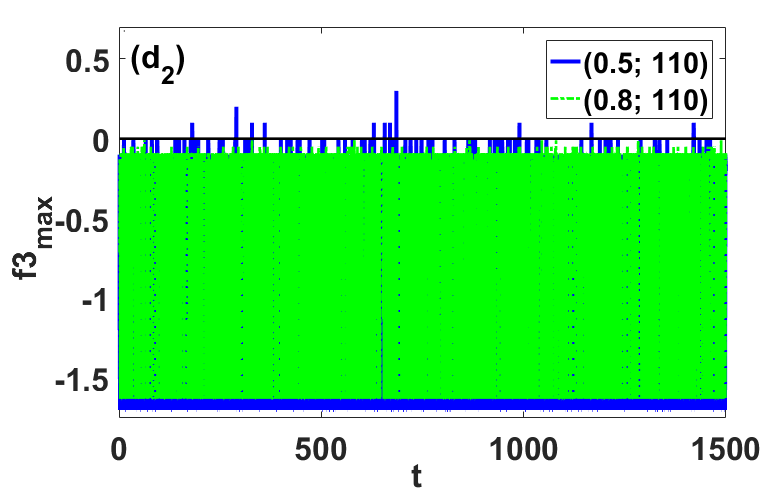}
    \caption{\footnotesize{The temporal evolution of the expressions defining the stability conditions for synchronization between the clusters for three pairs of values of $\epsilon$ and ${s_0}$. ($a_1$, $a_2$) illustrate the dynamics of $fe_{max}$, ($b_1$, $b_2$) show the dynamics of $f1_{max}$, ($c_1$, $c_2$) demonstrate that of $f2_{max}$ and ($d_1$, $d_2$) show the dynamics of $f3_{max}$. Other parameters: $\mu =1$, $v =5$, $u=2$, ${D_0}=2$, ${d_0}=2$.}
    } \label{fig1staba}
  \end{center}
\end{figure}

In contrast to the previous results, the pairs \((\epsilon=0.5, s_0=110)\) and \((\epsilon=0.8, s_0=110)\) show a convergence toward zero for \(f1_{\text{max}}\) and \(f2_{\text{max}}\) (see Figs.\ref{fig1staba}($b_2$) and ($c_2$)), with \(fe_{\text{max}} \leq 0\) (see Fig.\ref{fig1staba}($a_2$)) and  \(f3_{\text{max}} \leq 0\) (see Fig.\ref{fig1staba}($d_2$)).

\subsection{Mutual influence of vision ranges of the oscillators and the agents}\label{sec3c}

In this section, we focus our study on analyzing a model where there exists a bidirectional interaction between the internal dynamics (i.e., oscillators) and the external dynamics (i.e., agents) while considering the mobility of the clusters. In other words, the internal dynamics are influenced by the external dynamics through the vision size of the agents, and likewise, the external dynamics are influenced by the vision size of the oscillators. It is worth noting that this idea was previously introduced in ref.\cite{nguefoue2021network} but the authors did not take into account the mobility of the clusters. Thus, we define the external dynamics taking into account the effect of the internal dynamics through the vision size of the oscillators in the bounded 2-dimensional space by the following Eq.\ref{eq10}.

\begin{equation}\label{eq10}
 \left\{ \begin{array}{l}
 {\eta^{k} _i}\left( {t + \Delta t} \right) = {\eta^{k} _i}\left( t \right) +
{u^{k}_i}\sum\limits_j^N {{G^{k}_{ij}}\left( t \right)\cos \left( {{\theta^{k}_j}\left( t
\right) - {\theta^{k}_i}\left( t \right)} \right),}  \\
 {\xi^{k}_i}\left( {t + \Delta t} \right) = {\xi^{k}_i}\left( t \right) +
{u^{k}_i}\sum\limits_j^N {{G^{k}_{ij}}\left( t \right)\sin \left( {{\theta^{k}_j}\left( t
\right) - {\theta^{k}_i}\left( t \right)} \right).}  \\
 \end{array} \right.
\end{equation}
where $G_{ij}^{k}$ encodes the temporal structure of the network, representing the elements of the connectivity matrix ($G$) in cluster $k (k=1,2.)$. It is defined as follows:
\begin{equation}\label{eq11a}
 {G^{k}_{ij}}\left( t \right) = \left\{ \begin{array}{l}
 1\,\,\,\,\,\,\,\,if\,\,\,{D^{k}_{ij}}\left( t \right) \le {D^{k}_0} \\
 0\,\,\,\,\,\,\,otherwise \\
 \end{array} \right.
\end{equation}
with $D_{ij}^k$ representing a non-physical distance between the $i^{th}$ and $j^{th}$ oscillators in the same cluster. Its expression is given by Eq.\ref{Dij}, where $D_0^k$ ($\forall k,  D_0^k = D_0$) determines the maximum non-physical distance defined in the space of the oscillator variables ($x_i^{1}, x_i^{2}, x_i^{3}$) that separates two interacting oscillators.

\begin{equation}\label{Dij}
{D^{1}_{ij}}\left( t \right) = \sqrt {{{\left( {x^{1}_{ij}}\left( t \right)
\right)}^2} + {{\left( {x^{2}_{ij}}\left( t \right) \right)}^2} + {{\left(
{x^{3}_{ij}}\left( t \right) \right)}^2},}
 \end{equation}
 where $ {x^{1}_{ij}}={{\left( {{x^{1}_i}\left( t \right) - {x^{1}_j}\left( t \right)}
\right)}} $; $ {x^{2}_{ij}}={{\left( {{x^{2}_i}\left( t \right) - {x^{2}_j}\left( t \right)}
\right)}} $ and $ {x^{3}_{ij}}={{\left( {{x^{3}_i}\left( t \right) - {x^{3}_j}\left( t \right)} \right)}} $. And the  distance $D^{2}_{ij}$ in the second cluster is obtained by replacing the $x$ variables of the first cluster with the $y$ variables of the second cluster.\\

\begin{figure}[htp!]
  \begin{center}
    \includegraphics[scale=0.22]{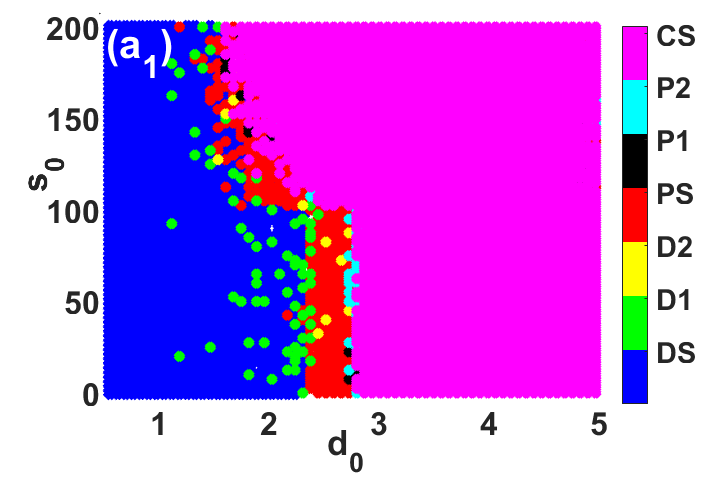}
    \includegraphics[scale=0.22]{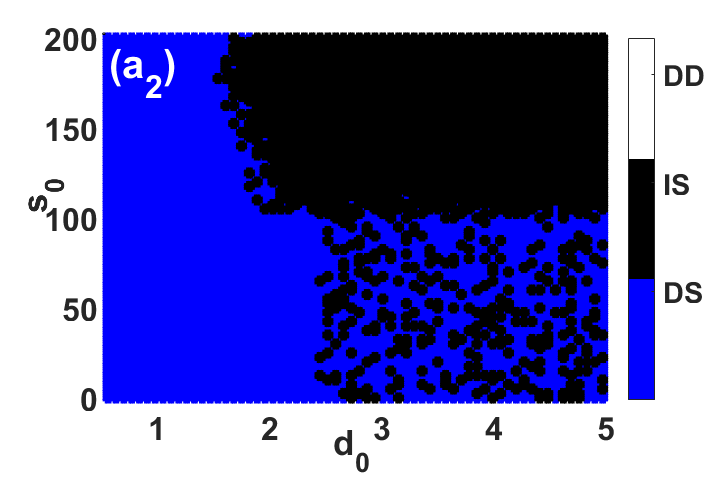}
    \includegraphics[scale=0.22]{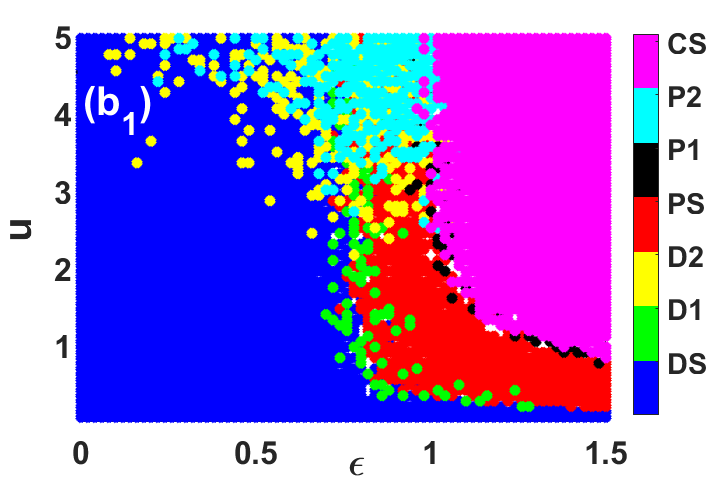}
    \includegraphics[scale=0.22]{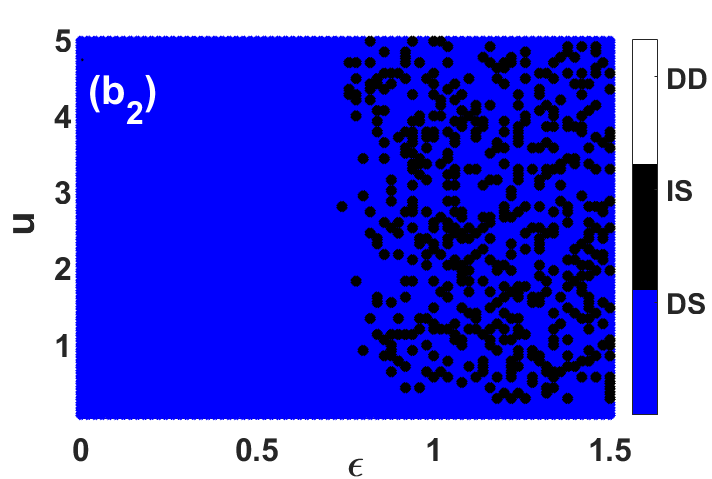}
    \includegraphics[scale=0.22]{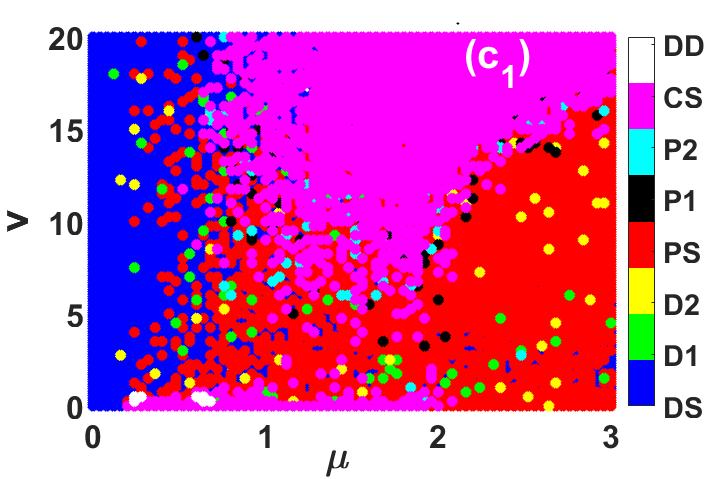}
    \includegraphics[scale=0.22]{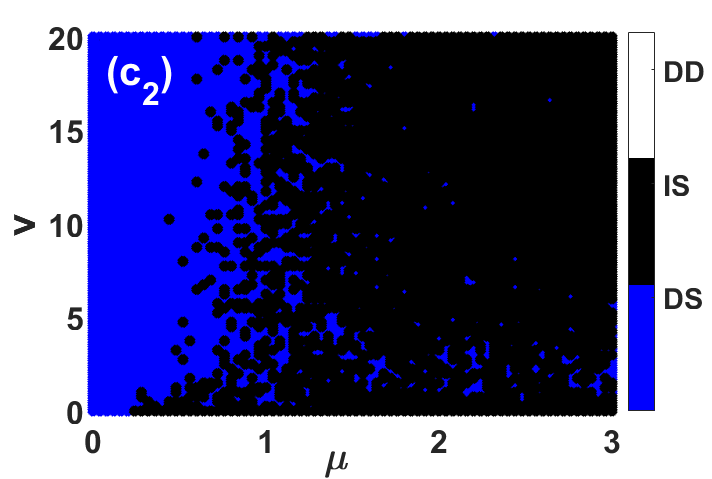}
    \caption{We report the different collective states achieved by the agents of the network, as a function of the following parameters: ($a_1, a_2$) $d_0$ and $s_0$ with $\mu=1$, $u=2$, ${D_0}=2$, $\epsilon=0.5$, $v =5$; ($b_1$, $b_2$) $\epsilon $ and $u$ with $\mu=1$, $v=5$, ${D_0}=2$, ${s_0}=100$, ${d_0}=2$; and  ($c_1$, $c_2$) $\mu$ and $v$ ($\epsilon=0.5$, $u=2$, ${D_0}=2$, ${s_0}=120$ and ${d_0}=2$). The first column corresponds to the dynamics within each cluster, and the second column corresponds to the dynamics between the clusters. The meanings of the different colors and acronyms are provided in Table.\ref{tab1}.
    }\label{acompila}
  \end{center}
\end{figure}

\begin{figure}[htp!]
  \begin{center}
    \includegraphics[scale=0.22]{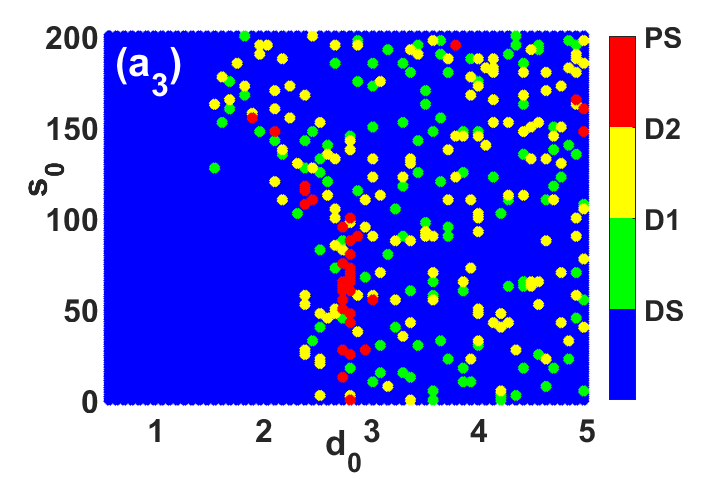}
    \includegraphics[scale=0.22]{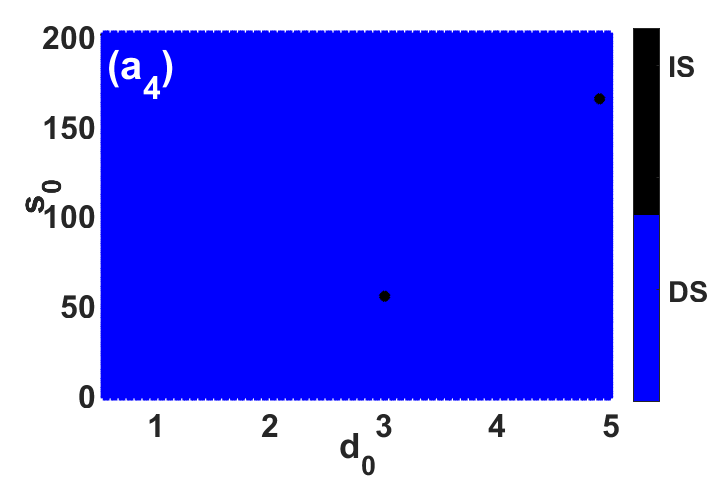}
    \includegraphics[scale=0.22]{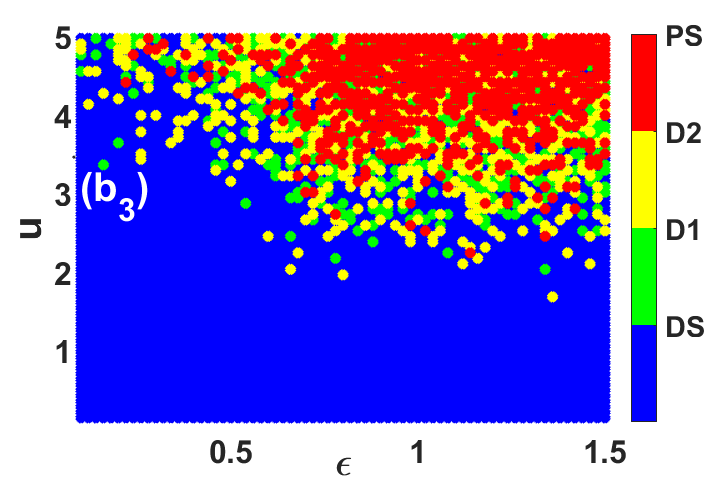}
    \includegraphics[scale=0.22]{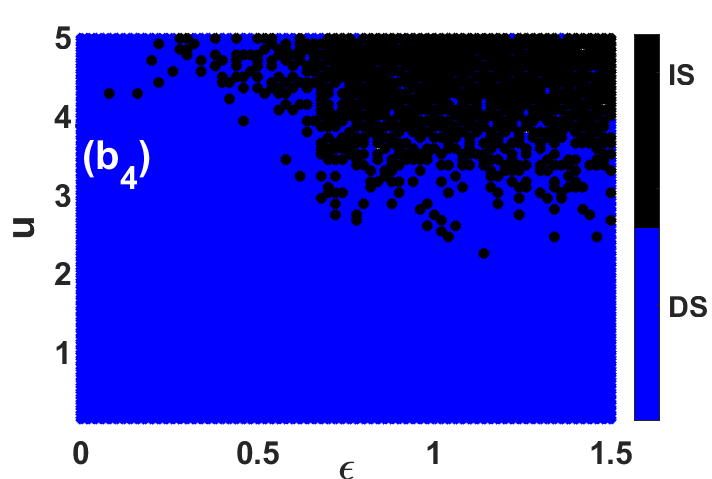}
    \includegraphics[scale=0.22]{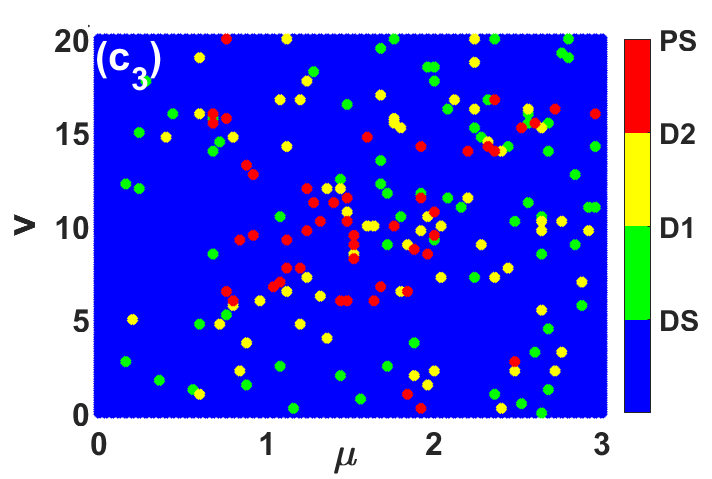}
    \includegraphics[scale=0.22]{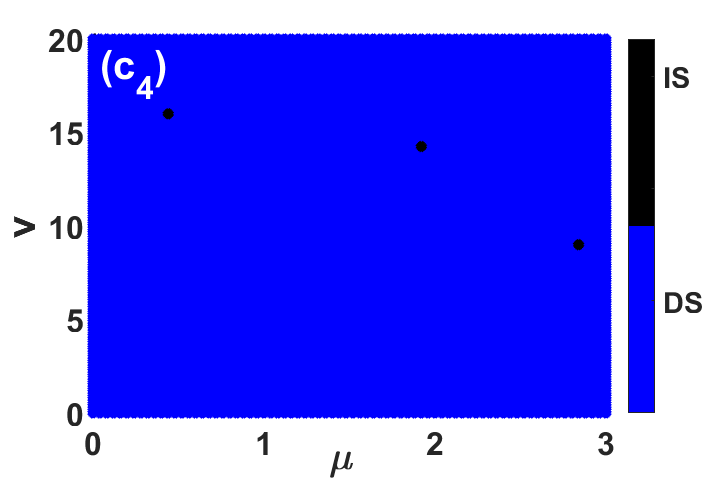}
    \caption{Presentation of the different collective behaviors formed by the agents in motion in the two-dimensional space as a function of: ($a_3, a_4$) $d_0$ and $s_0$ with $\mu=1$, $u=2$, ${D_0}=2$, $\epsilon=0.5$, $v =5$; ($b_3$, $b_4$) $\epsilon $ and $u$ with $\mu=1$, $v=5$, ${D_0}=2$, ${s_0}=100$, ${d_0}=2$; and  ($c_3$, $c_4$) $\mu$ and $v$ ($\epsilon=0.5$, $u=2$, ${D_0}=2$, ${s_0}=120$ and ${d_0}=2$). The first column corresponds to the dynamics within each cluster, and the second column corresponds to the dynamics between the clusters. The meanings of the different colors and acronyms are provided in Table.\ref{tab1}.} \label{acompilb}
  \end{center}
\end{figure}

The numerical analysis of Eqs.\ref{eq3}, \ref{eq3s}, and \ref{eq10} allows us to study the various collective behaviors exhibited by the entire system based on the control parameters. These collective behaviors are presented in Figs.\ref{acompila} (internal dynamics: oscillators) and \ref{acompilb} (external dynamics: agents), corresponding to the control parameter pairs ($d_0$, $s_0$), ($\epsilon$, $u$), and ($\mu$, $v$) defined in the previous section. In these figures, the colors and their meanings are exactly those used in the previous section.

Fig.\ref{acompila} depict the behavior of the internal dynamics within the clusters in the first column ($a_1$, $b_1$, and $c_1$), while the second column shows the behavior between the clusters ($a_2$, $b_2$, $c_2$). Under the influence of ${d_0}$ and ${s_0}$ (with other parameters fixed at $\mu=1$, $v=5$, ${D_0}=2$, $\epsilon=0.5$, $u=2$), Fig.\ref{acompila}($a_1$) displays several domains, among which the most dominant are the blue domain (DS) and the magenta domain (CS). These domains correspond to regions of desynchronized and completely synchronized states of the oscillators within each cluster. Between these regions, various other domains corresponding to different collective behaviors, illustrated by different colors (see Table I for description), are observed. Depending on the considered value of $s_0$, the transition from desynchronized state (DS) to complete synchronization (magenta) in the clusters passes through several intermediate collective states, represented by D1, D2, P1, P2, PS (see Table I for description). Comparing this result with the one in Fig.\ref{fig3}($a_1$), where the same investigation was conducted in the case of unidirectional influence between the internal and external parts of the system, the observed behaviors are quite similar, with the absence of the divergence region in the present analysis. The disappearance of the divergence zone indicates that the mutual interaction between oscillators and agents enables the internal dynamics to regulate and sustain oscillations for high values of the control parameters. For the same set of control parameters, Fig.\ref{acompila}($a_2$) illustrates the internal dynamics between the clusters. As previously suggested, for ${s_0}<100$, the occurrence of the black domain (i.e., synchronized domain) does not manifest continuously with the increase of the probability of intra-cluster connectivity. However, when $s_0>100$, as ${d_0}$ becomes progressively larger, achieving inter-cluster synchronization becomes almost inevitable.\\
Fig.\ref{acompilb}($a_3,a_4$) illustrates the results of the study of behaviors exhibited by agents in the space under the same conditions. The behaviors in the cluster depicted in Fig.\ref{acompilb}($a_3$) show an almost random evolution, with few instances of collective behaviors (synchronization phase) occurring in both clusters simultaneously. Moreover, the evolution of collective behavior formation within the clusters is challenging to control. This lack of collective behavior formation is evident in the dynamics between the clusters, as shown in Fig.\ref{acompilb}($a_4$), with almost a total absence of phase synchronization between the clusters. This confirms that achieving phase synchronization between the clusters requires synchronization within the clusters.

Considering the pair ($\epsilon,u$) as a control parameter, with the other parameters fixed such as $\mu=1$, $v=5$, ${D_0}=2$, ${s_0}=120$, and ${d_0}=2$, the transition from desynchronization to synchronization within the clusters also passes through various behaviors, as explained above. It is important to note here that at a fixed coupling strength, increasing the speed of agent movement helps collective behavior formations within the clusters (see Fig.\ref{acompila} ($b_1$)). The same observation is noted in the external dynamics (see Fig.\ref{acompilb}($b_3$)), and this velocity also contributes to the formation of collective behavior (in the agents) between the clusters (see Fig.\ref{acompilb}($b_4$)). Although the mutual influence between different parts of the system is taken into account, Fig.\ref{acompila}($b_2$) shows that it does not bring about significant improvement in achieving inter-cluster synchronization compared to what was obtained previously in Fig.\ref{fig3} ($b_2$).

Let's now investigate the dynamics considering the influence of the inter-cluster coupling strength $\mu$ and the cluster velocity $v$ (see Fig.\ref{acompila}($c_1,c_2$)) with the other parameters such as $\epsilon=0.5$, $u=2$, ${D_0}=2$, ${s_0}=120$, and ${d_0}=2$. Taking into account the mutual influence between the two parts of the system not only allows regulating the internal dynamics to limit any divergence but also shows a much faster transition (in terms of the required value of $\mu$) to achieve collective dynamics. This observation is also noted in the inter-cluster dynamics, as shown in Fig.\ref{acompila}($c_2$). The collective dynamics exhibited in space (see Fig.\ref{acompilb}($c_3,c_4$)) by the agents in this case remains very similar to that of Fig.\ref{acompila}($a_3,a_4$) commented above.

\section{Conclusion}\label{sec4}
Among the collective behaviors investigated in network science, synchronization stands out as one of the most extensively studied due to myriad multidisciplinary applications across technology, economics, biology, and social sciences. In this study, we explore the impact of cluster mobility on collective behaviors within a multi-cluster network of mobile oscillators. Our analysis reveals that synchronization between cluster is influenced by their spatial proximity. We primarily investigate two types of synchronization: complete synchronization observed exclusively within the oscillators and phase synchronization observed in the internal and external dynamics. The stability of complete synchronization is analytically and numerically demonstrated. These findings become even more intriguing when considering the occasional scenario where synchronization occurs in one cluster but not in both, despite the bidirectional interaction between clusters (see sec.\ref{sec3c}). The unique coupling model via the local center of mass within the clusters confers a wide range of applications to this work, such as in the study of social animals or herd behavior analysis, which could enhance understanding of their social dynamics, migration patterns, and ecological interactions. In the realm of technology, our research could be instrumental in studying drone swarms, offering versatile applications, including search and rescue missions by effectively coordinating drone fleets. Similarly, swarm robotic exploration could enable the deployment of robot teams to explore challenging environments such as disaster zones, space, or underwater areas, facilitating efficient exploration and data collection. Particularly in the technological domain, the mathematical analysis conducted could ensure stability and robustness during experimentation.

\section*{Acknowledgements}
 PL, TN, SJK, NMV and MoCLiS research group thanks ICTP for the equipment donation under the letter of donation Trieste $12^{th}$ August 2021. HAC acknowledges enlightening discussions with Albert Diaz Guilera. HAC and PL thank ICTP-SAIFR and FAPESP grant 2021/14335-0 for partial support. PL thanks the support of the Deutscher Akademischer Austausch Dienst (DAAD) at the Potsdam Institute for Climate Impact Research (PIK) under the Grant Number (91897150).  TN acknowledges support from the ``Reconstruction, Resilience and Recovery of Socio-Economic Networks'' RECON-NET EP\_FAIR\_005 - PE0000013 ``FAIR'' - PNRR M4C2 Investment 1.3, financed by the European Union – NextGenerationEU.


 	\appendix

\section{Influence of the number of nodes N }\label{appB}

\begin{figure}[htp]
  \begin{center}
  \begin{tabular}{ccc}
    \includegraphics[scale=0.18]{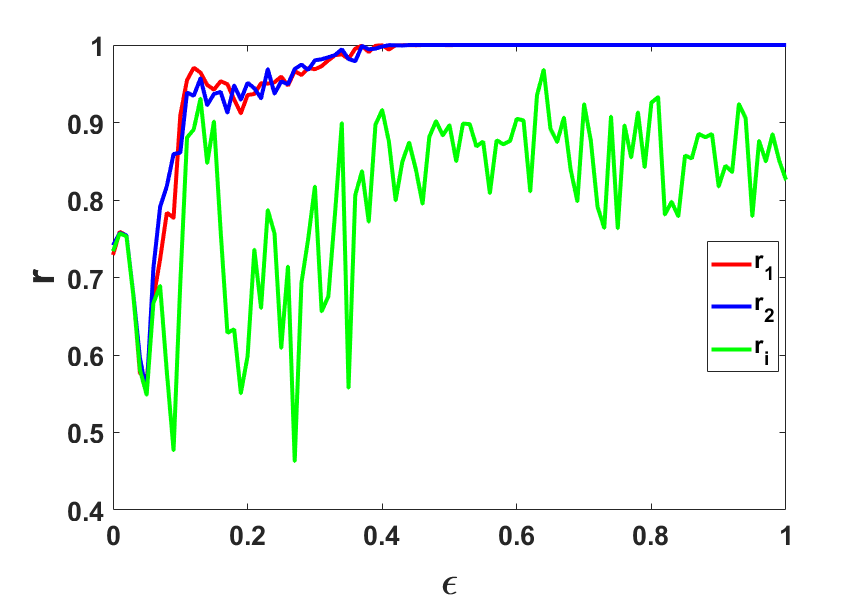}&
    \includegraphics[scale=0.18]{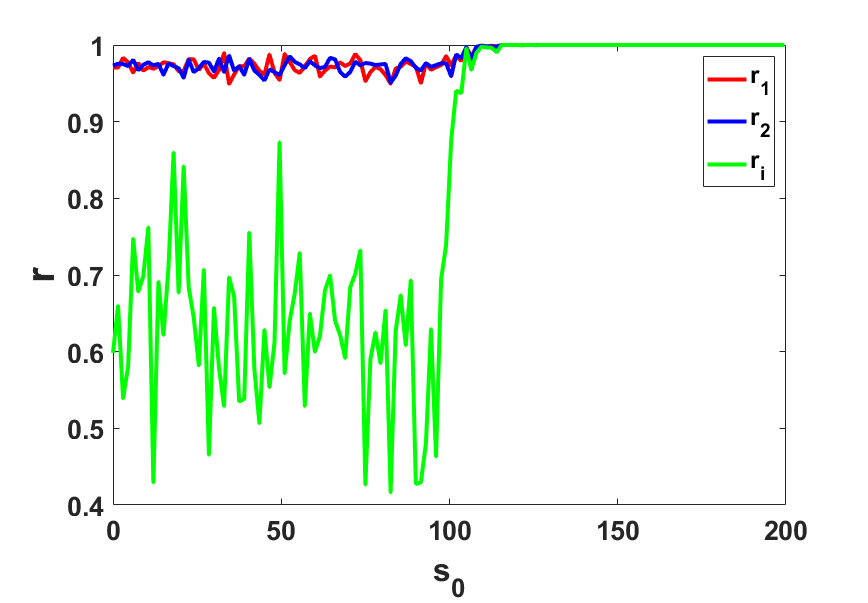}\\
    (a) & (b)
    \end{tabular}
    \begin{tabular}{cc}
    \includegraphics[scale=0.18]{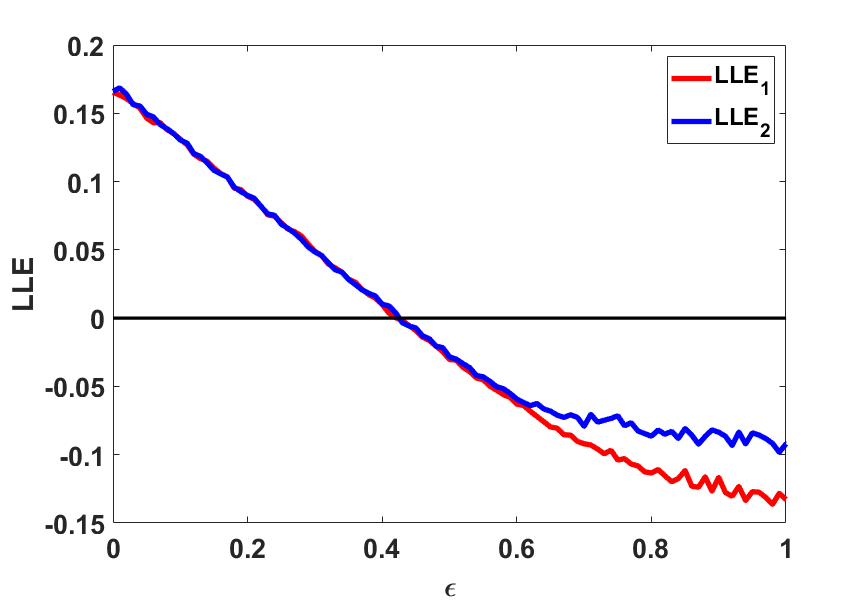}&
    \includegraphics[scale=0.18]{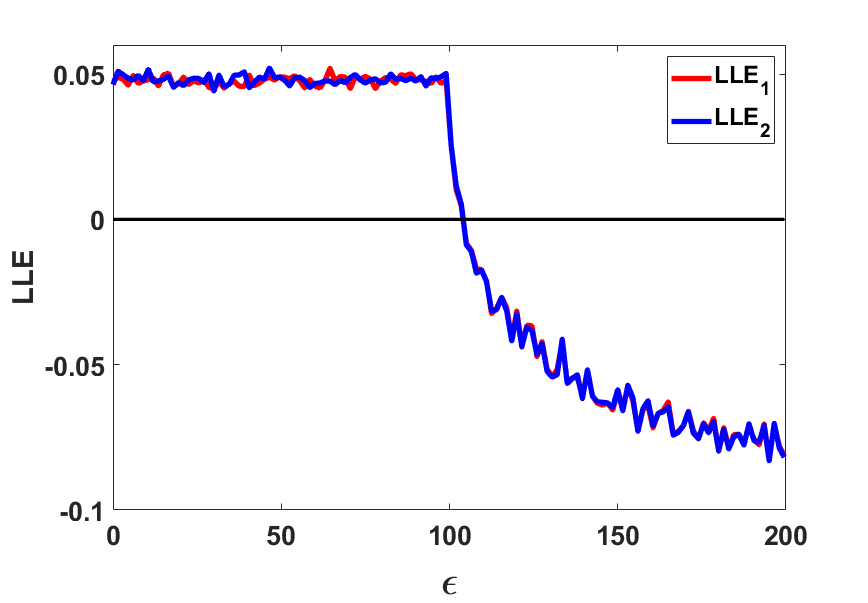}\\
    (c) & (d)
    \end{tabular}
    \caption{For \(N=100\), we illustrate the influence of intra-cluster coupling strength ($\epsilon$) and connectivity threshold between both clusters ($s_0$) on the network dynamics is investigated. By varying $\epsilon$. We show in (a) the order parameter depicting the collective dynamics of the phase of oscillators, and in (c) the Largest Lyapunov Exponent (LLE) showing the collective dynamics in terms of complete synchronization both within the same range of $\epsilon$ with $s_0=100$. Similarly, in (b) and (d), we analyze $s_0$ as the control parameter with $\epsilon = 0.35$. The red curve represents the dynamics of the first cluster, the blue curve represents the second cluster's dynamics, and the green curve illustrates the inter-clusterdynamics. Other parameter: $\mu = 1$, $u = 2$, $v = 10$, $d_{0} = 2$ .} \label{fig_rev_2}
  \end{center}
\end{figure}

\begin{figure}[htp]
  \begin{center}
  \begin{tabular}{ccc}
    \includegraphics[scale=0.18]{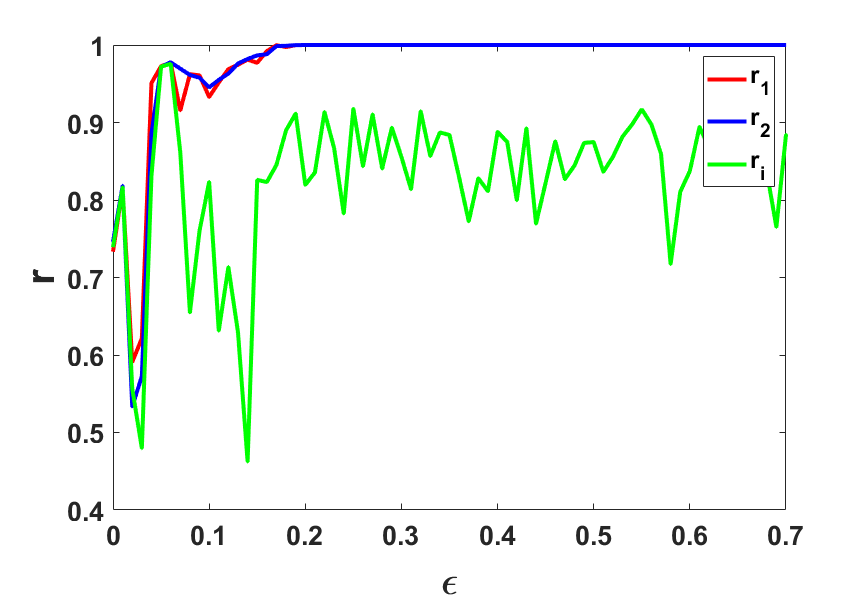}&
    \includegraphics[scale=0.18]{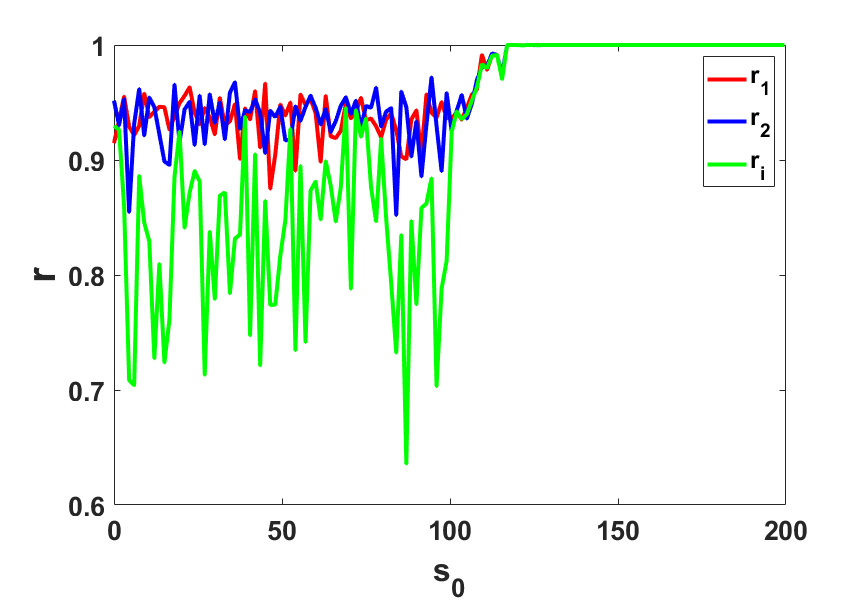}\\
    (a) & (b)
    \end{tabular}
    \begin{tabular}{cc}
    \includegraphics[scale=0.18]{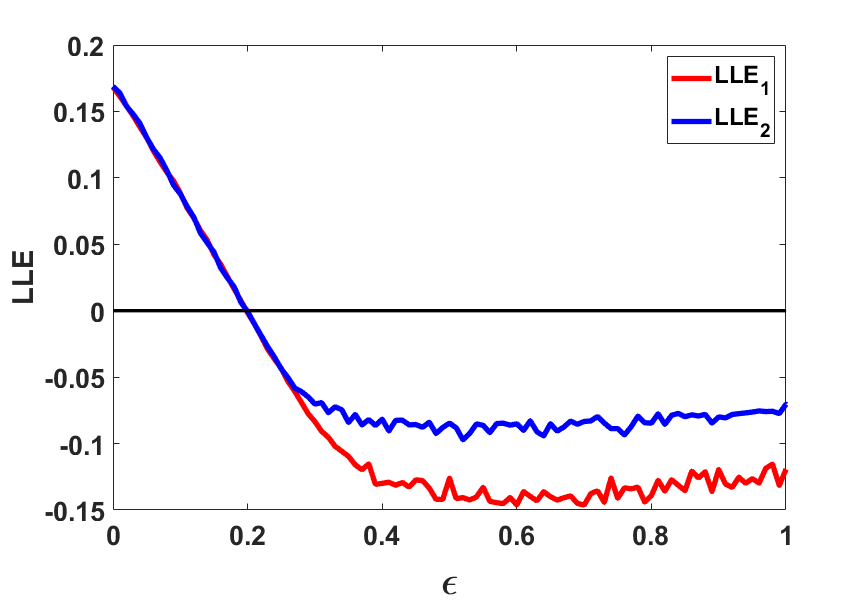}&
    \includegraphics[scale=0.18]{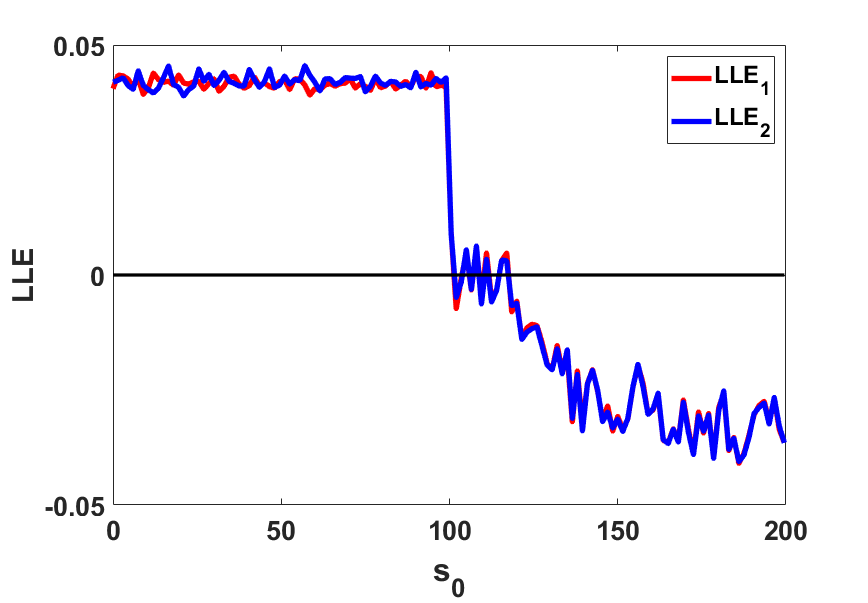}\\
    (c) & (d)
    \end{tabular}
    \caption{For \(N=200\), we illustrate the influence of intra-cluster coupling strength ($\epsilon$) and connectivity threshold between both clusters ($s_0$) on the network dynamics is investigated. By varying $\epsilon$. We show in (a) the order parameter depicting the collective dynamics of the phase of oscillators, and in (c) the Largest Lyapunov Exponent (LLE) showing the collective dynamics in terms of complete synchronization both within the same range of $\epsilon$ with $s_0=100$. Similarly, in (b) and (d), we analyze $s_0$ as the control parameter with $\epsilon = 0.15$. The red curve represents the dynamics of the first cluster, the blue curve represents the second cluster's dynamics, and the green curve illustrates the inter-clusterdynamics. Other parameter: $\mu = 1$, $u = 2$, $v = 10$, $d_{0} = 1.5$ .} \label{fig_rev_3}
  \end{center}
\end{figure}

Following the same methodology described in Section \ref{sec3}, we extended our analysis to two additional system sizes, namely $N = 100$ and $N = 200$. In this extension, the spatial dimensions of the $(Q \times Q)$ and $(P \times P)$ domains were kept constant. As a result, increasing the number of agents while maintaining fixed spatial dimensions leads to an increase in agent density. The results are shown in Figs.\ref{fig_rev_2} and \ref{fig_rev_3}, respectively. The first key observation is that all the behaviors seen with \(N=50\) are also observed for larger system size ($N=100$ and $N=200$), and the transition to stable states follows a similar pattern. The second observation is that as the size of the system increases, while keeping all other parameters constant, intra-cluster synchronization occurs more rapidly.
This can be attributed to the fact that, with the motion space dimension $(P \times P)$ remaining unchanged, each additional node increases the local density and enhances connectivity and information exchange among nodes. This, in turn, facilitates faster synchronization of the network’s dynamics. Conversely, the critical point at which behavioral changes occur remains relatively stable, depending on the vision size \(s_0\) between the clusters, since the number of clusters remains unchanged. In panels (b) and (d) of Figs.\ref{fig_rev_2} and Fig.\ref{fig_rev_3}, we set \(\epsilon = 0.35\) for Figs.\ref{fig_rev_2} and \(\epsilon = 0.15\) with \(d_0 = 1.5\) for Figs.\ref{fig_rev_3}, while keeping all other parameters constant.\\

	\nocite{*}
	\section*{References}
	\bibliography{ref}

\end{document}